\makeatletter
\newcommand{\algcaption}{%
\ifx \@captype \@undefined \@latex@error {\noexpand \caption outside float}\@ehd \expandafter \@gobble \else \refstepcounter \@captype \expandafter \@firstofone \fi {\@dblarg {\@caption \@captype }}%
}%
\makeatother

\documentclass[preprint]{iucr}              
\usepackage{color,xcolor}

\usepackage{xspace}
\RequirePackage{graphicx}
\usepackage{epstopdf}
\usepackage{mathtools, amsmath, amssymb, amsfonts}

\usepackage{algorithm}
\usepackage{algpseudocode}
\usepackage{amsmath}
\usepackage{graphics}
\usepackage{epsfig}

\usepackage[12pt]{moresize}
\usepackage{multirow}
\usepackage{enumitem}
\usepackage{url,ulem}
\usepackage{comment}
\usepackage{chemformula}

\newcommand{\be}{\begin{equation}}
\newcommand{\ee}{\end{equation}}

\newcommand{\eq}[1]{Eq.~\ref{eq:#1}}

\newcommand{\snmf}{{\sc stretchedNMF}\xspace}
\newcommand{\spsnmf}{{\sc sparse-stretchedNMF}\xspace}
\newcommand{\MMO}{{$\mathrm{MgMn_2O_4}$}\xspace}
\newcommand{\YOCl}{{$\mathrm{YOCl}$}\xspace}

\newcommand{\BTO}{{$\mathrm{BaTiO_3}$}\xspace}
\newcommand{\ZS}{{$\mathrm{ZnSe}$}\xspace}

\newcounter{saveenumi}

\makeatletter
\newcommand{\longdash}[1][2em]{%
\makebox[#1]{$\m@th\smash-\mkern-7mu\cleaders\hbox{$\mkern-2mu\smash-\mkern-2mu$}\hfill\mkern-7mu\smash-$}}

\newcommand{\bes}{\begin{enumerate}[wide, labelwidth=!, labelindent=0pt, label=\textbf{\textcolor{blue}{\arabic*}.}]}
\newcommand{\ees}{\end{enumerate}}

\definecolor{sobcolor}{HTML}{176d1b}

\definecolor{dgreen}{HTML}{008000}

\newcommand{\rw}{\ensuremath{R_w}\xspace}

\newcommand{\gr}{\ensuremath{G(r)}\xspace}

\newcommand{\iaa}{\AA\ensuremath{^{-1}}\xspace}

\graphicspath{{./figures/figs/}}

     \journalcode{M}              

\begin{document}                  



\title{Stretched Non-negative Matrix Factorization}
\shorttitle{}


\author[a]{Ran}{Gu}{}{}
\author[b]{Yevgeny}{Rakita}{}{}
\author[b]{Ling}{Lan}{}{}
\author[b]{Zach}{Thatcher}{}{}
\author[c]{Gabrielle~E.}{ Kamm}{}{}
\author[c]{Daniel}{O'Nolan}{}{}
\author[d]{Brennan}{Mcbride}{}{}
\author[d]{Allison}{Wustrow}{}{}
\author[d]{James~R.}{Neilson}{}{}
\author[c]{Karena~W.}{Chapman}{}{}
\cauthor[b]{Qiang}{Du}{qd2125@columbia.edu}{}
\cauthor[b]{Simon~J.~L.}{Billinge}{sb2896@columbia.edu}{}
\aff[a]{School of Statistics and Data Science, Nankai University, \city{Tianjin~300071}, \country{China}}
\aff[b]{Department of Applied Physics and Applied Mathematics, Fu Foundation School of Engineering \& Applied Sciences,
Columbia University, \city{New York, NY~10025} \country{USA}}
\aff[c]{Department of Chemistry, Stony Brook University, \city{Stony Brook, NY~11794}, \country{USA}}
\aff[d]{Department of Chemistry, Colorado State University, \city{
Fort Collins, CO~80523 }, \country{USA}}









\maketitle                        


\begin{abstract}
An algorithm is described and tested that carries out a non negative matrix factorization (NMF) ignoring any stretching of the signal along the axis of the independent variable.
This extended NMF model is called \snmf .
Variability in a set of signals due to this stretching is then ignored in the decomposition.
This can be used, for example, to study sets of powder diffraction data collected at different
temperatures where the materials are undergoing thermal expansion.
It gives a more meaningful decomposition in this case where the component signals resemble signals from chemical components in the sample.
The \snmf  model introduces a new variable, the stretching factor, to describe any expansion of the signal.
To solve \snmf, we discretize it and employ Block Coordinate Descent framework algorithms.
The initial experimental results indicate that \snmf model outperforms the conventional NMF
for sets of data with such an expansion.
A further enhancement to \snmf for the case of powder diffraction data from crystalline materials called \spsnmf, which makes use of the sparsity of the powder diffraction signals, allows correct extractions even for very small stretches where \snmf struggles.
As well as  demonstrating the model performance on simulated PXRD patterns and atomic pair distribution functions (PDFs), it also proved successful when applied to real data taken from an \textit{in situ} chemical reaction experiment.
\end{abstract}


\section{Introduction}\label{sec:1}



Nonnegative matrix factorization (NMF) is an unsupervised machine learning method used for decomposing compressed data.
NMF extracts distinct components from related signal sets in various research fields, including signal processing~\cite{buciuNonnegativeMatrixFactorization2008}, biomedical engineering~\cite{sraNonnegativeMatrixApproximation2006}, pattern recognition~\cite{cichockiFastLocalAlgorithms2009}, image engineering~\cite{buciuNonnegativeMatrixFactorization2008a} and so on. NMF differs from principle component analysis (PCA) \cite{jolliffePrincipalComponentAnalysis2002} by applying positivity constraints on the extracted components and their weights. It is then attractive for attempting to find components that resemble physical signals in the case where the positivity constraints are expected to hold.  In crystallography, NMF has demonstrated significant potential in finding physically plausible structural signals from diffraction data collected from \textit{in situ} chemical reactions~\cite{longRapidIdentificationStructural2009a, kusneHighthroughputDeterminationStructural2015a, huaNonequilibriumMetalOxides2021}. Recently, NMF has also been used for \textit{in situ} time-dependent diffraction measurements~\cite{liuValidationNonnegativeMatrix2021a,thatc;aca22} and spatially resolved electron diffraction maps~\cite{rakit;am23}, single-layer nanosheets~\cite{beauvaisResolvingSinglelayerNanosheets2021}, integrated multimodal analysis~\cite{onolanMultimodalAnalyticalToolkit2021}, and metal–organic frameworks~\cite{chenNodeDistortionTunable2023,rayderUnveilingUnexpectedModulatorCO22023}.

The conventional NMF method assumes that the components remain fixed with respect to time, and therefore can hardly capture changes in the components over time. For example, in temperature series experiments, increased temperature can expand the inter-atomic distances, resulting in the stretching of peak positions in the measured powder diffraction pattern or atomic pair distribution function (PDF) data.

To address this limitation, extended NMF models have been proposed. One such model is the Shifted NMF, which accounts for shifts in the onset of a frequency profile, which can be induced by the Doppler effect for spectrometry data~\cite{morupShiftedNonNegativeMatrix2007}. However, Shifted NMF is not able to solve temperature series data problem because the change in the component is a stretch, not a shift. Another approach is to incorporate stretching regression steps into the analysis workflow~\cite{rakitaActiveReactionControl2020c}.

In this paper, we propose a new extended NMF model called \snmf, to explore a more fundamental aspect of the algorithm itself. We introduce a stretching factor matrix to describe the stretching scales of each component and each component is allowed to have different entire stretching factors at different moments. \snmf can be developed to account for a simple stretching of the measured signal and returns only components that explain variability beyond this stretching.

In this paper, we first develop the mathematical formulas of \snmf in the form of functional optimization.  We present the method of discretization and the optimization algorithm. Then using both simulated and real data, we show that \snmf significantly outperforms conventional NMF in the case of diffraction data with thermal expansion. Furthermore, we show that the algorithm may be used to extract different chemical components from the data if there are multiple components that have differential thermal expansivities. This gives an interesting possibility for extracting the components in a multi-phase sample from a temperature dependent measurement of that sample, even when those components are not changing chemically during the measurement. Although we focused on diffraction signals from temperature series data, the algorithm may be used for any case where part of the changes to the signal are exactly, or approximately, a stretch of its dependent variable.

\section{Stretched Nonnegative matrix factorization}\label{sec:2}

Nonnegative matrix factorization (NMF) is a mathematical tool to approximate a given matrix  $Z\in \mathbb{R}^{N\times M}$  by the product of two low-rank nonnegative matrices,
\begin{equation}
Z\approx XY,
\end{equation}
where $X\in \mathbb{R}^{N\times K}$ and $Y\in \mathbb{R}^{K\times M}$, and $K\ll N,M$~\cite{leeLearningPartsObjects1999d}. Its description and use is described in detail in multiple places~\cite{berryAlgorithmsApplicationsApproximate2007b,wangNonnegativeMatrixFactorization2013a}.
The common NMF model uses the square of Euclidian distance (SED) as the objective function, and the corresponding optimization problem is written as
\begin{align}\label{prob:nmf}
\min\limits_{X\in \mathbb{R}^{N\times K},Y\in \mathbb{R}^{K\times M}} & \quad\frac{1}{2} \left\|XY-Z\right\|_F^2, \nonumber\\
\text{s.t. }\hspace{10mm}  & \quad X\geq 0\text{ and }Y\geq 0.
\end{align}

Similar to principal component analysis~\cite{abdiPrincipalComponentAnalysis2010a}
, the NMF decomposition will find components that explain variability in the signals in the set of data.
Unlike PCA, a constraint of positivity is applied to both the components and the weights.  Since many real physical signals, and their weights, obey positivity, NMF is more likely to find components that resemble signals from different physical components contributing to a compound signal coming from multiple sources.  As such, it is finding extensive use in scientific applications~\cite{renNonnegativeMatrixFactorization2018,liuValidationNonnegativeMatrix2021a,gobinetApplicationNonnegativeMatrix2004}.

Here we address a situation where one aspect of the variability, a stretching of the signal on the axis of its independent variable, is not of scientific interest, for example, due to the thermal expansion of a material affecting its diffraction pattern.   We formulate an approach named \snmf which extends the conventional NMF decomposition whilst accounting for the stretching in the algorithm.

Suppose the experimental signals, which are columns of $Z$, $z^m$ for $m=1\dots M$, and the components, which are columns in $X$, $x_k$ for $k=1\dots K$, are continuous functions of an independent variable $r$. Then the conventional NMF optimization problem may be written as
\begin{equation}\label{eq:continuousnmf}
\min\limits_{y_k^m\geq 0,x_k\geq 0} \quad \sum\limits_{m=1}^M \left\|\sum\limits_{k=1}^K y_k^m x_k(r)-z^m(r)   \right\|_{L_2}^2,
\end{equation}
where $y_k^m$ is the weight of the $k$th component at the $m$th position in the dataset. Now, we assume that there is an $m$-dependent stretching of the signal along the $r$ axis. The component signals stretch with component-dependent rates that we capture in a stretching factor, $\{a_k^m\}_{m=1,\dots,M}$.
We add the stretching factors $a_k^m$ into \eq{continuousnmf} and the optimization problem becomes
\begin{equation}\label{eq:contistrechednmf}
\min\limits_{a_k^m,y_k^m\geq 0,x_k\geq 0} \quad\sum\limits_{m=1}^M \left\|\sum\limits_{k=1}^K y_k^m x_k(r/a_k^m)-z^m(r)  \right \|_{L_2}^2.
\end{equation}
Notice that if $a_k^m>1$, $x_k$ is stretched, and if $a_k^m<1$, $x_k$ is compressed. In practice, we consider a finite $r$ range $[0,r_{\max}]$. Therefore, without loss of generality, we define $x_k(r)=0$ for $r\geq r_{\max }$. Thus, when $a_k^m>1$, $x_k(r/a_k^m)=0$ for $r\geq r_{\max }/a_k^m$. Now we are able to expand the $L_2$ norm in equation~(\ref{eq:contistrechednmf}) as an integral over the $r$ range as
\begin{align}\label{eq:contistrechednmfint}
\min\limits_{a_k^m,y_k^m\geq 0,x_k\geq 0} & \quad\sum\limits_{m=1}^M \int_0^{r_{\max}}\left(\sum\limits_{k=1}^K y_k^m x_k(r/a_k^m )-z^m(r)\right)^2 dr, \nonumber\\
\text{s.t. }\hspace{6mm}  & \quad x_k(r)=0\text{, if } r\geq r_{\max }.
\end{align}

For fixed component $k$, $\{a_k^m\}_{m=1,\cdots,M}$ is a series of stretching factors, which usually change smoothly with time $m$. However, the optimization problem in equation~(\ref{eq:contistrechednmfint}) is non-convex, and hence the smoothness of $\{a_k^m\}_{m=1,\cdots,M}$ may be violated when we solve it numerically. Therefore, we add a regularization term to the objective function to make it favor smooth $a_k$, i.e.,
\begin{align}\label{eq:smoothcontinuenmf}
\min\limits_{a_k^m,y_k^m\geq 0,x_k\geq 0} & \quad \sum\limits_{m=1}^M \int_0^{r_{\max}}\left(\sum\limits_{k=1}^K y_k^m x_k(r/a_k^m )-z^m(r)\right)^2 dr \nonumber\\
& \hspace{30mm} + \rho\sum\limits_{k=1}^K\sum\limits_{m=1}^{M-2}(a_k^m-2a_k^{m+1}+a_k^{m+2})^2, \nonumber\\
\text{s.t. }\hspace{6mm}  & \quad x_k(r)=0\text{, if } r\geq r_{\max },
\end{align}
where $\sum\limits_{k=1}^K\sum\limits_{m=1}^{M-2}(a_k^m-2a_k^{m+1}+a_k^{m+2})^2$ is the smoothness regularization and $\rho$ is the parameter to control the effect of regularization. In our numerical testing section, we initiate a large $\rho$ and gradually decrease it in subsequent iterations.

\section{Numerical Solution of \snmf}\label{sec:3}

In this section, we describe the numerical implementation of the \snmf.

In order to numerically solve the functional optimization problem~(\ref{eq:smoothcontinuenmf}), we discretize the functionals and solve the corresponding vector optimization problem. Unlike Shifted NMF~\cite{morupShiftedNonNegativeMatrix2007}, we cannot get benefits from discretizing the frequency domain of the components after applying the Fourier transform. So we choose to discretize the problem in the real $r$ space, without loss of generality, using a uniform grid on $[0,r_{\max}]$. Since we have introduced the stretching factors, when we discretize the functionals $x_k(r/a_k^m)$, on this uniform grid the arguments $r/a_k^m$ are actually not on the grid nodes. Therefore, we apply a spline interpolation, that is we approximate $x_k(r/a_k^m)$ from $x_k(r)$, where the interpolant is a piecewise polynomial. In terms of the order of the spline, we need at least a quadratic order, i.e., a piecewise quadratic polynomial with continuous derivatives on the grid points. The smoothness of the spline will help the convergence of the discretized optimization solution. In this paper, we use a quadratic spline interpolation to approximate $x_k(r/a_k^m)$ in the optimization problem~(\ref{eq:smoothcontinuenmf}).  Explicitly, let $0=r_0<r_1<\dots<r_{n}=r_{max}$ be the uniform grid nodes, resulting in an interval of $h=r_{max}/n$. The quadratic piecewise polynomial approximation, $S_i(r)$, of $x(r)$ for $r\in [r_i,r_{i+1}]$ is
\begin{align}
	S_i(r)=q_i(r-r_i)(r-r_{i+1})+\left[x(r_{i+1})-x(r_i)\right](r-r_i)/h+x(r_i),
\end{align}
where $q_i$ is the quadratic coefficient to be determined. The derivatives of the polynomials $S_i(r)$ and $S_{i+1}(r)$ are
\begin{align}
	&S_i'(r)=q_i(2r-r_i-r_{i+1})+\left[x(r_{i+1})-x(r_i)\right]/h, \label{eq:polyderivetive1}\\
	&S_{i+1}'(r)=q_{i+1}(2r-r_{i+1}-r_{i+2})+\left[x(r_{i+2})-x(r_{i+1})\right]/h. \label{eq:polyderivetive2}
\end{align}
Notice the fact that the second-order spline should have continuous derivatives over the entire domain, which means that $S_{i}'(r_{i+1})=S_{i+1}'(r_{i+1})$ at positions $r_{i+1}$ for $i=0,\dots,n-2$,
using~(\ref{eq:polyderivetive1}) and~(\ref{eq:polyderivetive2}), we get
\begin{align}
	q_i+q_{i+1} = [x(r_i) - 2x(r_{i+1}) + x(r_{i+2})]/h^2.
\end{align}
Since we have $x(r)=0$, for $r\geq r_{max}$, we set $S_{n-1}(r_n)=0$ and $S'_{n-1}(r_n)=0$. Then we can write $q$ as
\begin{equation}\label{eq:qandx}
\left(
         \begin{array}{c}
           q_0 \\
           q_1 \\
           \vdots \\
           q_{n-1} \\
         \end{array}
       \right)=\frac{1}{h^2}\left(
  \begin{array}{lllll}
    1 & 1 &  &  &  \\
     & 1 & 1 &  &  \\
     &  & \ddots &  \ddots &  \\
     &  &  & 1 & 1 \\
     &  &  &  & 1 \\
  \end{array}
\right)^{-1}\left(
                 \begin{array}{rrrrr}
                   1 & -2 & 1 &  &  \\
                    & 1 & -2 & 1 &  \\
                    &  &  \ddots& \ddots & \ddots \\
                    &  &  & 1 & -2 \\
                    &  &  &  & 1 \\
                 \end{array}
               \right)\left(
                        \begin{array}{c}
                          x(r_0) \\
                          x(r_1) \\
                          \vdots \\
                          x(r_{n-1}) \\
                        \end{array}
                      \right).
\end{equation}

Now we can write $x_k(r/a_k^m)$ in terms of $x_k(r_i)$ as a linear transformation
\begin{equation}\label{eq:x}
	x_k(r/a_k^m)=q_i(r/a_k^m-r_i)(r/a_k^m-r_{i+1})+[x(r_{i+1})-x(r_{i})](r/a_k^m-r_i)/h+x(r_{i}),
\end{equation}
if $r/a_k^m\in [r_i,r_{i+1}]$ and $x_k(r/a_k^m)$ is set to zero if $r/a_k^m\geq r_{max}$. Since the leading coefficient $q$ is also linearly dependent on $x$ as shown in equation~(\ref{eq:qandx}), we can conclude the linear transformation $x_k(r_i/a_k^m) = b_{i,a_k^m}^Tx_k$ for $i=0,1,\dots,n$ in a matrix form
\begin{equation}
\left(
  \begin{array}{c}
    x_k(r_0/a_k^m) \\
    x_k(r_1/a_k^m) \\
    \vdots \\
    x_k(r_n/a_k^m) \\
  \end{array}
\right)
=\left(
   \begin{array}{c}
     \longdash \hspace{2mm} b_{0,a_k^m}^T \hspace{2mm}\longdash\\
     \longdash \hspace{2mm} b_{1,a_k^m}^T \hspace{2mm}\longdash\\
     \vdots \\
     \longdash \hspace{2mm} b_{n,a_k^m}^T \hspace{2mm}\longdash\\
   \end{array}
 \right)
 \left(\begin{array}{c}
   x_k(r_0) \\
   x_k(r_1) \\
   \vdots \\
   x_k(r_n) \\
 \end{array}\right),
\end{equation}
and denote the coefficient matrix as $B_{a_k^m}$. Now we are ready to write the discretization of the optimization problem in~(\ref{eq:smoothcontinuenmf}) as
\begin{equation}\label{eq:finalnmf}
\min\limits_{a_k^m,\,y_k^m,\,x_k\geq 0} \sum\limits_{m=1}^M \left\|\sum\limits_{k=1}^K y_k^m B_{a_k^m} x_k-z^m \right\|^2+\rho\sum\limits_{k=1}^K\sum\limits_{m=1}^{M-2}(a_k^m-2a_k^{m+1}+a_k^{m+2})^2,
\end{equation}
where $y_k^m$, $x_k$ and $z^m$ are discretized functionals on the uniform grid $0=r_{0}<r_{1}<\cdots<r_{n}=r_{\max}$.

If the theoretical convergence is neglected, linear interpolation may be used as an approximation. In this case, we set $q_i=0$ in~(\ref{eq:x}). The final form of the optimization problem is still~(\ref{eq:finalnmf}), but with a different $B_{a_k^m}$ with higher sparsity.

Among the existing methods, a popular approach to solve the conventional NMF is alternating non-negative least squares (ANLS)~\cite{paateroPositiveMatrixFactorization1994,linProjectedGradientMethods2007,kimNonnegativeMatrixFactorization2008,guanNeNMFOptimalGradient2012,huangQuadraticRegularizationProjected2015}. ANLS alternatively adjusts $X$ and $Y$ to minimize the objective function and each subproblem can be solved by the non-negative linear least square method. In fact, this framework is also called the block coordinate descent (BCD) method with two blocks. In our problem \eq{finalnmf}, which can be simplified as
\begin{equation}\label{eq:finalsimp}
\min\limits_{A,Y,X\geq 0} f(A,Y,X),
\end{equation}
there are three blocks $A$, $Y$ and $X$. Applying the BCD method with three blocks, we can solve the problem~\eq{finalsimp} using algorithm~\ref{alg:0}.
\begin{algorithm}
\algcaption{Block Coordinate Descent (BCD) Method}
\label{alg:0}
\begin{algorithmic}[1]
\For{$t=1,2,\cdots$}
\State $A:=\arg\min_{A\geq 0} \langle \nabla_A f(\hat A,Y,X),A\rangle $ \label{algo:sub_A}
\State $Y:=\arg\min_{Y\geq 0} \langle \nabla_Y f(A,\hat Y,X),Y\rangle $ \label{algo:sub_Y}
\State $X:=\arg\min_{X\geq 0} \langle \nabla_X f(A,Y,\hat X),X\rangle $ \label{algo:sub_X}
\EndFor
\end{algorithmic}
\end{algorithm}

Similar to conventional NMF, the subproblems of $Y$ and $X$ in Lines~\ref{algo:sub_Y} and~\ref{algo:sub_X} are convex quadratic programming problems that can be easily solved by existing solvers. But the subproblem of $A$ in Line~\ref{algo:sub_A} is highly non-convex and therefore we cannot efficiently solve it for its global minimum. In practice, we use a subspace trust-region method~\cite{colemanInteriorTrustRegion1996} to find a local minimum.

The convergence of the BCD method for 3~blocks is not guaranteed~\cite{grippoConvergenceBlockNonlinear2000}. Here we use an algorithm that can guarantee its convergence for a quadratic spline approximation that is called the linearized block coordinate descent method~\cite{xuGloballyConvergentAlgorithm2017}. The outline of the framework is presented in Algorithm~\ref{alg:1}, where $\alpha_t$ is the step size and $\hat A/\hat X/\hat Y$ are the extrapolations of the current $A/X/Y$ and previous $A/X/Y$. In each iteration, the algorithm randomly chooses one block and minimizes the corresponding linear approximation and a proximal term. One can refer to \cite{xuGloballyConvergentAlgorithm2017} for more information about parameter selections.
\begin{algorithm}
\algcaption{linearized block coordinate descent method}
\label{alg:1}
\begin{algorithmic}[1]
\For{$t=1,2,\cdots$}
\State pick one of the following to implement in a deterministic or random manner;
\State $A:=\arg\min_{A\geq 0} \langle \nabla_A f(\hat A,Y,X),A\rangle + \frac{1}{\alpha_t}\|A-\hat A\|^2$
\State $Y:=\arg\min_{Y\geq 0} \langle \nabla_Y f(A,\hat Y,X),Y\rangle + \frac{1}{\alpha_t}\|Y-\hat Y\|^2$
\State $X:=\arg\min_{X\geq 0} \langle \nabla_X f(A,Y,\hat X),X\rangle + \frac{1}{\alpha_t}\|X-\hat X\|^2$
\EndFor
\end{algorithmic}
\end{algorithm}

\section{Diffraction Use Case}

\subsection{Introduction}

Here we test the approach using simulated and also real x-ray powder diffraction (PXRD) data~\cite{pecharskyFundamentalsPowderDiffraction2008, dinnebierPowderDiffractionTheory2008a}, and atomic pair distribution function (PDF)~\cite{egamiBraggPeaksStructural2012d} data. PXRD and PDF patterns are continuous 1D signals that encode the 3D arrangement of atoms in a material. We assume a situation where the PDF and PXRD patterns have been measured for samples as a function of temperature and are undergoing thermal expansion, where the thermal expansion coefficient of each sample is different. The thermal expansion causes Bragg peaks in the PXRD, and peaks in the PDF, to change their positions. In principle, thermal expansion can be different along different directions of the crystal, but often it is quite isotropic and appears as a stretching of the pattern where the peak shifts increase with increasing distance along the independent variable axis as required for this algorithm to work. This makes it an interesting use-case for \snmf, though we note that the \snmf\ may be applied to any series of signals where one aspect of the variability is a continuous stretching on the axis of the independent variable.

The goal of our testing use-case is to see if we can use NMF in general, but \snmf in particular, to separate the chemical components in a binary chemical mixture where the two components have different thermal expansion coefficients. For example, this could be used by a material scientist to discover the chemical components in a synthesis product by measuring the mixture as a function of temperature and running \snmf on the mixture, where the algorithm returns mathematical components that resemble the PXRD or PDF signals of the actual chemical components. These mathematical components could then be given to algorithms such as the {\sc structureMining}~\cite{yangStructureminingScreeningStructure2020g} or {\sc spacegroupMining}~\cite{liuUsingMachineLearning2019} algorithms that are implemented as a service on the PDFitc.com website~\cite{yangCloudPlatformAtomic2021c}. These algorithms, given an uploaded PDF, will return a rank ordered list of candidate structures consistent with that PDF.  Our test of the algorithm will therefore consist of taking either simulated or actual measured data over a wide temperature range from binary mixtures where the components have different thermal expansion coefficients.  These signals will be fed to NMF and \snmf to extract two components which will then be analyzed to see if they resemble the signals from the actual chemical components.  In the case of the PDF, an interesting test of this is to take the extracted mathematical components and giving them to the {\sc structureMining} algorithm to see if it correctly identifies the chemical component from the \snmf and conventional NMF extraction.

\subsection{Data}

To evaluate the performance of the \snmf and \spsnmf algorithms, we tested them on the following test-case datasets:
\subsubsection{Simulated PXRD and PDF data with increasing lattice parameters}
We used a set of 20 simulated PXRD and PDF patterns. The PXRD and PDF were from a weighted sum of a simulated cubic perovskite \BTO phase  and a cubic wurtzite \ZS. The \BTO:\ZS phase-fraction was set to 1:1, which corresponded to an atomic concentration ratio 0.61:1.00 when initial lattice parameters of \BTO=4.18\AA\ and \ZS=5.62\AA\ were used. The CIF files used for \BTO and \ZS were from structures reported in \cite{keler_reaction_1960} and~\cite{andreev_synthesis_1995}, respectively, and downloaded from the Springer Materials database (\url{https://materials.springer.com/isp/crystallographic/docs/sd\_0304044}, \url{https://materials.springer.com/isp/crystallographic/docs/sd\_1929775}).

To simulate lattice expansion we assumed a constant expansion coefficient, $\alpha_{BTO}$ and $\alpha_{ZS}$, for each component. For a more complete exploration of the performance of conventional NMF with that of \snmf and \spsnmf, we did not restrict ourselves to expansion coefficients that resemble actual thermal expansivities. However, since the linear thermal expansion coefficient of \BTO is approximately twice of that of \ZS~\cite{bland_thermal_1959, su_thermal_2009}, we always set the expansion coefficient of \BTO to be twice that of \ZS.

We note that to test purely the effects of stretching, which is the basis of the current NMF modification, we fixed and did not vary the atomic displacement parameters (ADPs) that would result in changes in the attenuation of the PXRD Bragg peaks and broadening of peaks in the simulated PDF.
Such effects are likely to be present in real data and may require a further modification to the NMF algorithm in the future but this is beyond the scope of the current paper.
This set of simulations assumed no phase transition or chemical reaction to be occurring and the relative weights of the components were not varied in the computed dataset.

The PXRD patterns were simulated using Dans-Diffraction \cite{porter_danporterdans_diffraction_2020}.
Pseudo-Voight lineshapes were used.
The PDFs were simulated using Diffpy-CMI \cite{juhas_complex_2015}.
The code used to generate the PDFs is reproduced in the supplementary information and can be found at \url{https://github.com/yevgenyr/diffpysim}.
The static set of parameters used for the simulations is reproduced in Table~\ref{tab:SIMparam}.
 \begin{table}[]
    \centering
    \caption{The static set of parameters that were used for PXRD and PDF simulations.}
    \begin{center}
    \begin{tabular}{|c|c|c|c|}
    \hline
 					& Parameter &	Value & Units	\\
\hline
\multirow{8}{*}{PDF}	&steps&	20 & \\
               				&qmin&	0.1 & $\mathrm{\AA^{-1}}$\\
              				&qmax&	30 &  $\mathrm{\AA^{-1}}$\\
               				&qdamp&	 0.03 & \\
               				&Uiso&	0.007 & $\mathrm{\AA^{2}}$\\
               				&rmin&	0& \\
               				&rmax&	120&  $\mathrm{\AA}$\\
               				&rstep&	0.01& $\mathrm{\AA}$\\
\hline
\multirow{4}{*}{PXRD}	&voigt\_profile.sigma&	1.5& \\
               				&voigt\_profile.gamma&	1.5& \\
              				&qmax&	30& $\mathrm{\AA^{-1}}$\\
               				&Uiso&	0.007& $\mathrm{\AA^{2}}$\\

\hline
    \end{tabular}
    \end{center}
    \label{tab:SIMparam}
\end{table}

\begin{figure}
\includegraphics[scale=0.5]{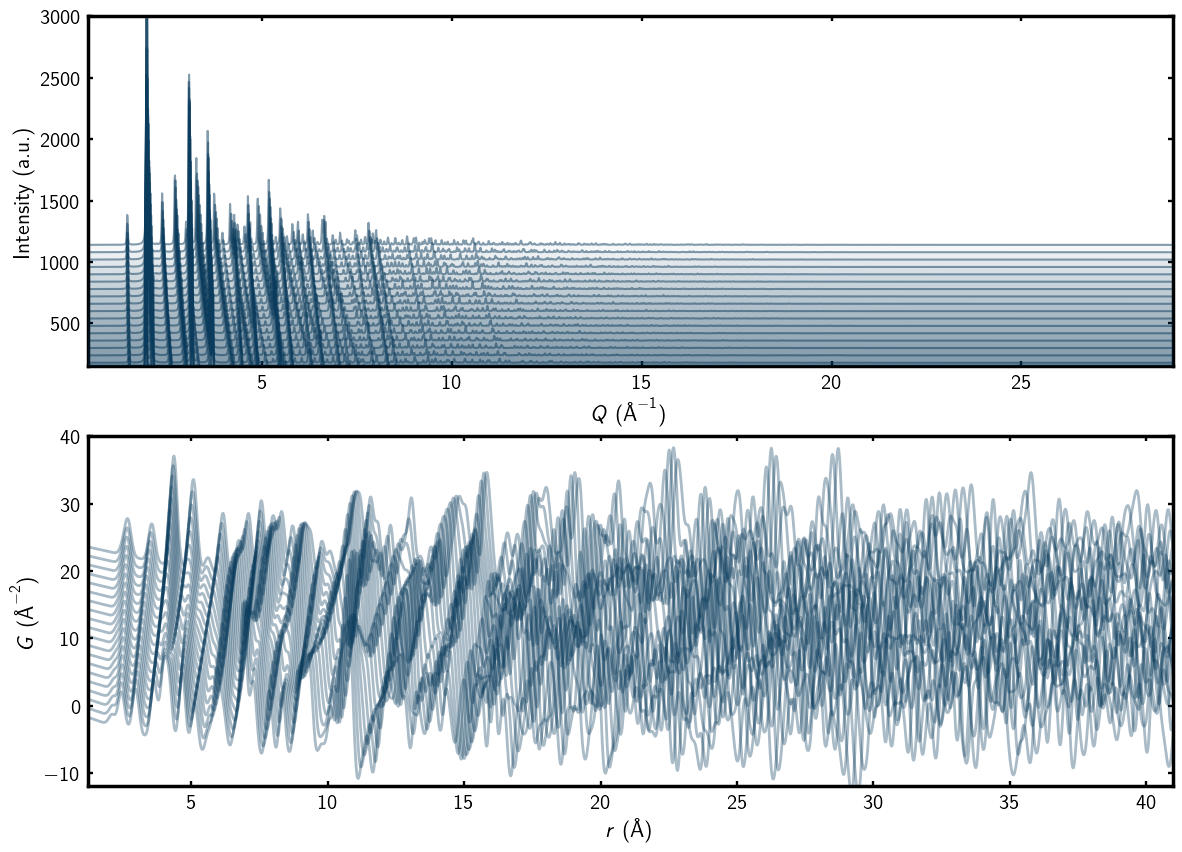}
  	\caption{Example simulated signals used in the tests. These were from a linear combination of the PDFs of \ch{BaTiO3} and \ch{ZnSe} where each was computed with a linearly expanding cubic lattice parameter. The PDFs are plotted offset from one another expanding from the bottom to the top of the figure. In the curves shown an overall expansion values of 20\% and 10\% was used from the first to the last curve. Top panel shows the simulated XRD and the bottom panel shows the PDF.}
   \label{simu_input_strain10}
\end{figure}

Representative PXRD and PDF patterns are shown in Fig.~\ref{simu_input_strain10}.

\subsubsection{Experimental PXRD data - thermal expansion}\label{sec:dataEXP1}
To test the \snmf and \spsnmf algorithms on real data, we use part of an \textit{in situ} solid-state synthesis reaction dataset where no phase-transition or chemical reaction occurred but which spanned a rather broad temperature range.
This allows us to evaluate how the algorithms perform for the effect of thermal expansion of a phase mixture from real data.
The PXRD experiment was done at the 28-ID-2 beamline (XPD instrument) at the NSLS-II facility at Brookhaven National Laboratory.
A large area 2D Perkin Elmer detector was used to acquire the data.
To gain high spectral resolution in the PXRD,  the distance between the sample and the detector was set to 144~cm. The beam wavelength was 0.1949~\AA.

A stoichiometric mixture of 2:1 \YOCl ($>$98\% tetragonal phase) and \MMO (spinel phase) was uniformly mixed and sealed in a quartz capillary.
It was then heated in a gradient furnance, meaning that each location on the quartz tube had a different temperature.\cite{onolan_thermal-gradient_2020}
The absolute temperatures at each point along the sample were calibrated from the lattice expansion of a known calibration material, Ni.
The data went from a low temperature of 368$\mathrm{^oC}$ to a highest temperature of 668$\mathrm{^oC}$ with a total of 20 individual temperature points. Using `pyFAI'~\cite{ashiotis_fast_2015}, the collected 2D diffraction patterns were then cleaned by masking the beam-stop and over-bright/dead pixels, followed by an azimuthal integration to gain 1D PXRD patterns.
The 1D PXRD data was then used as inputs to the different NMF algorithms.

\subsubsection{Experimental PXRD data - thermal expansion and reaction}
We also tested the NMF algorithms on another PXRD dataset, but this time, where a solid-state chemical reaction happened together with the thermal expansion so that the weights of the components as well as the thermal expansion were varying during the experiment.
The data were measured as the temperature changed from 28~C to 370~C in 215 steps during the reaction of
$$\ch{CuCl2} + \ch{Na2Se2}\longrightarrow \ch{CuSe2}+2\ch{NaCl}. $$
Here the components involved in the reaction are NaCl, CuSe, \ch{Cu2Se}, Se, pyrite, and marcasite, as determined by a multi-phase Rietveld refinement on the full dataset carried out previously.
The full details of experiment are published in \cite{martinolichPolymorphSelectivitySuperconducting2015}.

\subsection{Algorithm Developments}

\begin{figure}
  	\begin{center}
  	\includegraphics[scale=0.6]{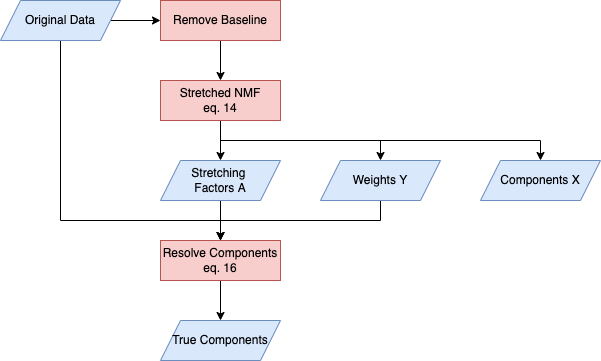}
  	\caption{The \snmf\ workflow of the PDF test.}
  	\label{fig:workflow_pdf}
  	\end{center}
\end{figure}

    In the case of PDF data, we apply \snmf to time-series data according to the workflow shown in the chart in Fig.~\ref{fig:workflow_pdf}.
      A common experimental function (for example, the output of xPDFsuite \cite{yang;arxiv14} and PDFgetX3 \cite{juhas2013pdfgetx3}, is the \gr function \cite{farrowRelationshipAtomicPair2009a}.  This function oscillates above and below zero. NMF works on the basis that signals are positive and in order to avoid the loss of signal where the function goes negative, we need to modify the signal into a non-negative form. Here we use an offset method, by taking the smallest of all data values and adding its absolute value to all data. This approach has the advantage of being simple and has been successfully applied to the deep learning method of PDF analysis \cite{liuValidationNonnegativeMatrix2021a}.

After running the NMF solvers, we must restore the components to valid $G(r)$ functions (oscillating around zero).
To do this we utilize the solved weights and stretching factors to recover the components according to
\begin{equation}\label{eq:nmf2pdf}
\min\limits_{x_k} \sum\limits_{m=1}^M \left\|\sum\limits_{k=1}^K y_k^m B_{a_k^m} x_k-z^m \right\|^2,
\end{equation}
where, $z^m$ is the original data rather than the data after the offset pre-processing and the other symbols are described alongsided Eq.~\ref{eq:finalnmf}.  The weight,~$y$,  and stretching factors,~$a$, are fixed to be those obtained from the NMF solution, and we remove the constraint that the components must be non-negative. Functions resembling $G(r)$ are then recovered from the NMF components and may be fit using standard PDF modeling protocols. This is reasonable because it is based on our trust in the weights and stretching factors of the NMF solver's solution of the preprocessed data. This approach is highly automated and can be applied to both conventional NMF and \snmf, because the stretching factor of the conventional NMF is always $1$.

For the case of PXRD data from highly crystalline samples, we have the additional observation that the spectrum consists of a sparse set of sharp peaks.  That is, the function value is zero in between the Bragg peaks (neglecting backgrounds and any diffuse scattering). We can make use of this property to enhance our ability to decompose signals by adding a sparse regularization term to the optimization problem. For the case where there are smooth backgrounds in experimental PXRD data, the background can be easily and automatically eliminated to make the data sparse. In this case we make two modifications to the optimization problem in Eq.~\eqref{eq:finalnmf}.
The first is adding the $l_{1/2}$ sparse regularization term to $x$ \cite{xuL1Regularization2010}. The second is adding an upper bound on $y$, in order to prevent $x$ from collapsing to zero as a whole, resulting in
\begin{align}\label{eq:sparsestretchnmf}
\min\limits_{a_k^m>0,\,0\leq y_k^m\leq 1,\,x_k\geq 0}
& \sum\limits_{m=1}^M \left\|\sum\limits_{k=1}^K y_k^m B_{a_k^m} x_k-z^m \right\|^2 + \\\nonumber
& \rho\sum\limits_{k=1}^K\sum\limits_{m=1}^{M-2}(a_k^m-2a_k^{m+1}+a_k^{m+2})^2 +\eta \sum\limits_{k=1}^K \sum\limits_{i=1}^n (x_{k,i})^{\frac{1}{2}} .
\end{align}
We refer to this as \spsnmf.

In this optimization model, there are two regularization parameters, $\rho$ and $\eta$. From experience, the smoothness parameter $\rho$ is not sensitive and is usually adjusted by multiplying by 10. The sparsity parameter $\eta$ can be adjusted by doubling.

\section{Numerical Results}

\subsection{Results on Simulated PDF}

First, we compare the performance of the conventional NMF and the \snmf on simulated PDF data. The PDFs were generated by a combination of two components, namely simulated \ch{BaTiO3} and ZnSe. The weight coefficients for each component were set as constants. We also assigned different linearly increasing rates for the thermal expansion of \ch{BaTiO3} and ZnSe. Specifically, we used artificially generated rates such that \ch{BaTiO3} and \ch{ZnSe} linearly expands from the first PDF to the last with 20\% and 10\% expansions, respectively.

We then applied the conventional NMF and \snmf methods to extract two components from the simulated PDF data. These could then be compared with the ground-truth PDFs. In principal, any of the ground-truth PDFs could be picked as we apply a stretching factor to the NMF component signal before the comparison. In this study, we selected the ground-truth PDF that resulted in the minimal residual when using only the scale factor variable. We further optimized the agreement between the NMF component and the selected ground-truth PDF by varying both the scale-factor and stretch factor variables.

\begin{figure}
\includegraphics[scale=0.6]{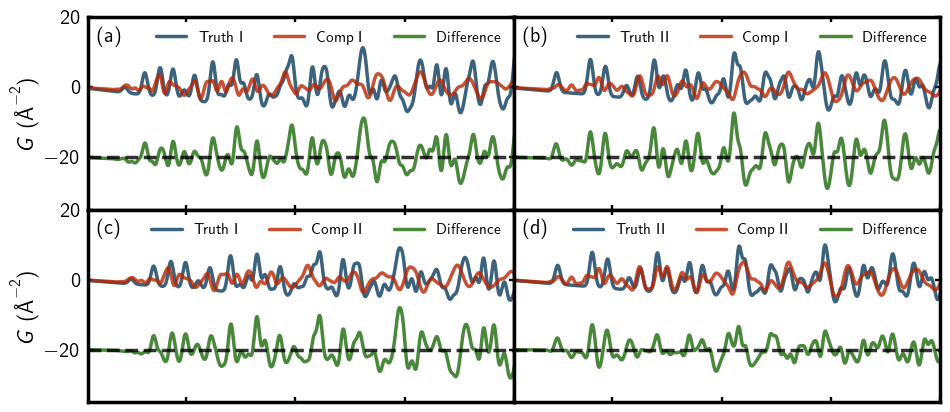}
\hspace{1.5mm}\includegraphics[scale=0.6]{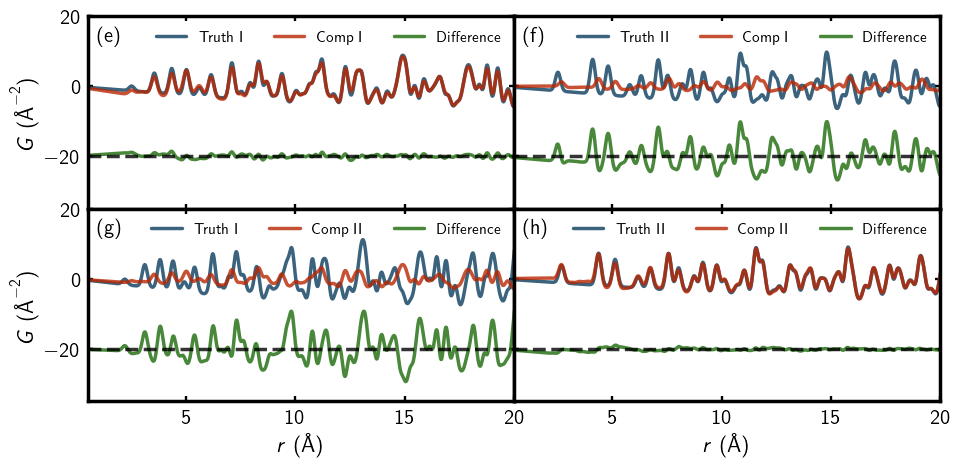}
  	\caption{Comparison of NMF-extracted PDF signals (red) and ground truth PDFs (blue).
  The top $2\times 2$ block (a-d) shows the extraction for the conventional NMF algorithm and the
  bottom block (e-h) is the extraction for the \snmf algorithm.
  The exact curves are indicated by the legend in each case, where Truth I and Truth II are the simulated ground truth curves I and II and Comp I and II indicate the first and second extracted components in each case.}  	\label{fig:simuPDF_nmf10}
\end{figure}

We first evaluated the outcomes of the conventional NMF approach. These findings are illustrated in Fig.~\ref{fig:simuPDF_nmf10}(a-d) and Tab.~\ref{tab:simuPDF10}.
 \begin{table}[]
    \centering
    \caption{Results of the comparison between the NMF extracted components and the ground-truth PDFs on simulated PDF data test. \rw and PC are the residual and the Pearson Correlation values between the numerical solution and the ground truth.}
    \begin{center}
    \begin{tabular}{|c|c|c|c|}
    \hline
Method & Component &	Ground Truth	& $R_w$ (PC)\\
\hline
\multirow{4}{*}{Conventional NMF}&1&\multirow{2}{*}{\ch{BaTiO3}}&0.9488 (0.3157) \\
                &2&      &\pmb{0.9050 (0.4255)}\\
                \cline{2-4}
                &1&\multirow{2}{*}{ZnSe}	 &0.9186 (0.3953)\\
                &2&		 &\pmb{0.8462 (0.5328)}\\
\hline
\multirow{4}{*}{\snmf}   &1&\multirow{2}{*}{\ch{BaTiO3}}&\pmb{0.1357 (0.9911)}\\	
                &2&		 &0.9750 (0.2225)\\
                \cline{2-4}
                &1&\multirow{2}{*}{ZnSe}  &0.8600 (0.5104)\\
                &2&		 &\pmb{0.1162 (0.9937)}\\
                \hline
    \end{tabular}
    \end{center}
    \label{tab:simuPDF10}
\end{table}
Fig.~\ref{fig:simuPDF_nmf10}(a-d) depicts the resulting PDFs in a matrix layout, with the NMF extracted components being represented as rows (in red) and the ground-truth PDFs as columns (in blue).
The difference curves (ground-truth - NMF component) are plotted below in green.
Large residuals and large \rw factors are evident between all the NMF components and the ground-truth curves and the NMF extraction has failed to produce components that resemble the actual signals.
This is not surprising since the weights of the two components are not varying in the test.

The same test was applied using the \snmf algorithm and the results are shown in Fig.~\ref{fig:simuPDF_nmf10}(e-h) and Tab.~\ref{tab:simuPDF10}.
In this case we can see that the \snmf extracted signal I is closely related to the ground truth component I and likewise for the component II.
This is evident as a very flat difference curve in Fig.~\ref{fig:simuPDF_nmf10}(e) and (h) and small \rw for these pairings in Tab.~\ref{tab:simuPDF10}.

This shows that even in the absence of changes in component weights the \snmf algorithm can extract components just from a differential stretching of the structure signal.

\subsection{Results on Simulated PXRD}

We carry out the same comparison of NMF vs \snmf for the case of powder diffraction signals.
Similar to the simulated PDF case, the data comprise of a combination of simulated \ch{BaTiO3} and ZnSe, where \ch{BaTiO3} and ZnSe have 20\% and 10\% linearly varying expansions, respectively.

\begin{figure}
  	\includegraphics[scale=0.67]{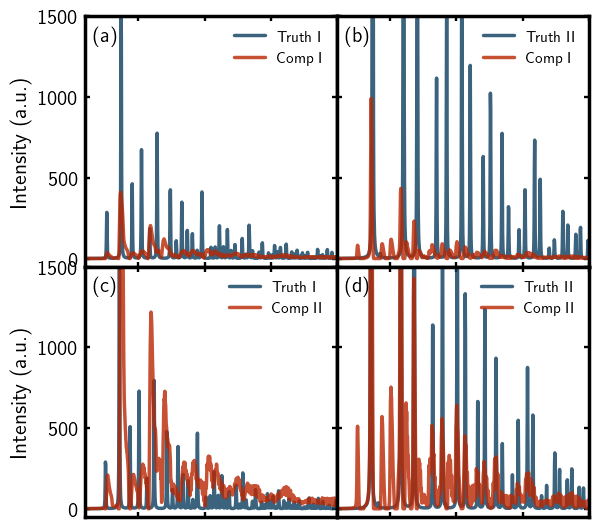}\\
	\hspace{5mm}\includegraphics[scale=0.67]{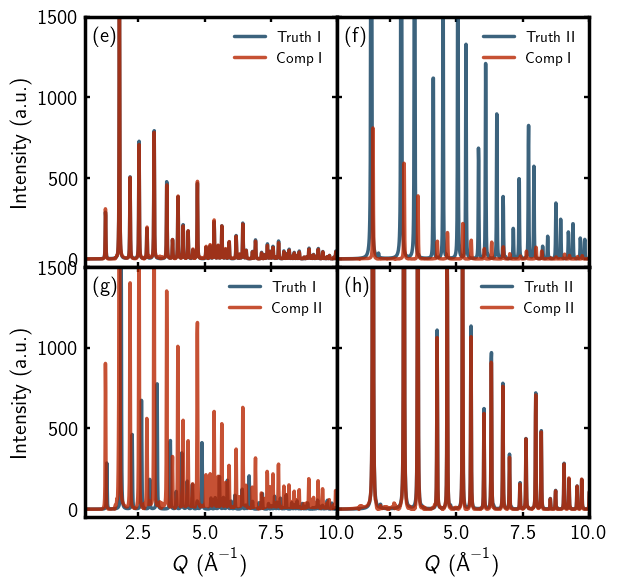}
  	\caption{NMF and \snmf solutions on simulated PXRD data. Truth I is the simulated powder diffraction pattern for \ch{BaTiO3} and Truth II is for ZnSe. The blue and red curves represent the true PDF and the extracted NMF components, respectively. (a)-(d) show the comparisons between components extracted with conventional NMF and (e)-(h) those extracted using the \snmf algorithm.}
  	\label{fig:simuXRD_strnmf_strain10}
\end{figure}

The results of the comparison are presented in Fig.~\ref{fig:simuXRD_strnmf_strain10} and Tab.~\ref{tab:simuXRD10}.

\begin{table}[]
    \centering
    \caption{Comparison between the NMF extracted components and the ground-truth PXRDs on the simulated PXRD data set. \rw and PC are the residual and the Pearson Correlation values, respectively, between the numerical solutions and the ground truth. }
    \begin{tabular}{|c|c|c|c|}
    \hline
Method & Component &	Ground Truth	& $R_w$ (PC) \\
\hline
\multirow{4}{*}{Conventional NMF}&1&\multirow{2}{*}{\ch{BaTiO3}}&0.8506 (0.5040)\\
                &2&      & \pmb{0.7196 (0.6830)}\\
                \cline{2-4}
                &1&\multirow{2}{*}{ZnSe}	 &0.7976 (0.5979)\\
                &2&		 &\pmb{0.5655 (0.8225)}\\
\hline
\multirow{4}{*}{\snmf}   &1&\multirow{2}{*}{\ch{BaTiO3}}&\pmb{0.0437 (0.9990)}\\	
                &2&		 &0.8427 (0.5290)\\
                \cline{2-4}
                &1&\multirow{2}{*}{ZnSe}  &0.8400 (0.5326)\\
                &2&		 &\pmb{0.0272 (0.9996)}\\
                \hline
    \end{tabular}

    \label{tab:simuXRD10}
\end{table}
As is evident in Fig.~\ref{fig:simuXRD_strnmf_strain10}(a-d), none of the extracted conventional NMF components resemble ground-truth curves.  Again, this is not a surprise because the weights of the components are not changing.
However, for the \snmf extraction we see that the first extracted component (Comp I) corresponds well the \ch{BaTiO3} pattern (Truth I), and the second extracted component (Comp II) corresponds well to the \ch{ZnSe} diffraction pattern (Truth II) (Fig.~\ref{fig:simuXRD_strnmf_strain10}(e) and (h)).

As with the simulated PDF data, the \snmf algorithm can extract components resembling the physical signals from a phase mixture where the weights are not changing but there is a variable thermal expansion.

\subsection{Results on simulated PDF and PXRD data with small expansion coefficients}

The tests above show that even in the presence of large stretches of signals \snmf can automatically extract signals that resemble real physical signals whereas conventional NMF cannot, at least in the case where the component weights are not changing.

We now would like to see how well \snmf can perform for smaller stretching factors, for example, for magnitudes that might occur in physical systems due to thermal expansion.
The simulated data is still taken as the combination of \ch{BaTiO3}\ and \ch{ZnSe}.
The weights are set to constants as before.
However, in this example we set the thermal expansion rates of \ch{BaTiO3}\ and \ch{ZnSe} to 4\% and 2\%, respectively.
Both simulated PDF and PXRD are tested.

\begin{figure}
  	\begin{center}
  \includegraphics[scale=0.55]{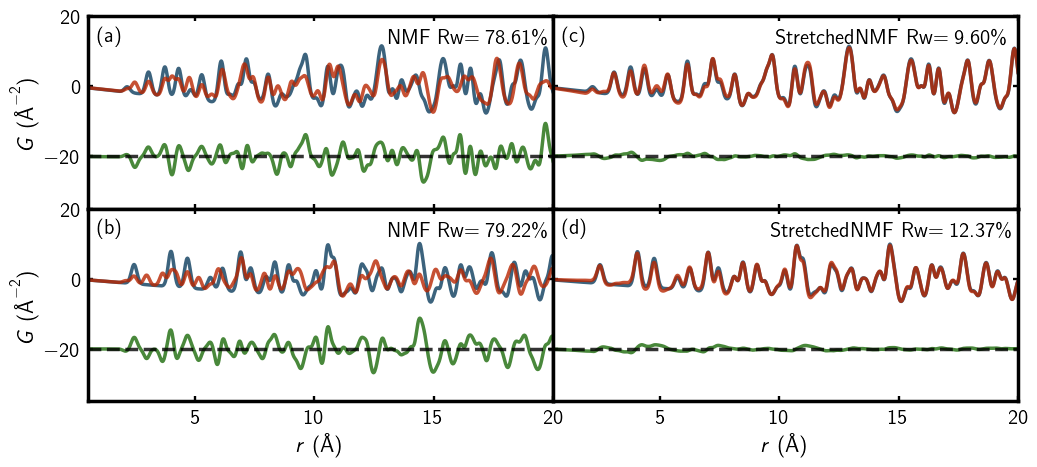}
  	\caption{Extracted components compared to the ground-truth components for PDFs computed from the structures of \ch{BaTiO3} (blue curve in (a) and (c)) and ZnSe (blue curve in (b) and (d)). Extracted signals done with conventional NMF (red curves in (a) and (b)) and \snmf (red curves in (c) and (d)).}
  	\label{fig:simuPDF_compare_error}
  	\end{center}
\end{figure}

First, we compare the performance of the conventional NMF and the \snmf on simulated PDF data. The results are presented in Tab.~\ref{tab:simuPDF} and Fig.~\ref{fig:simuPDF_compare_error}.
\begin{table}[]
    \centering
    \caption{Results of the comparison between the NMF extracted components and the ground-truth PDFs on simulated PDF data sets with 2\% expansions on ZnSe. \rw and PC are the residual and the Pearson Correlation values between the numerical solution and the ground truth.}
    \begin{center}
    \begin{tabular}{|c|c|c|c|}
    \hline
Method & Componenpt &	Ground Truth	& $R_w$ (PC)\\
\hline
\multirow{4}{*}{Conventional NMF}&1&\multirow{2}{*}{\ch{BaTiO3}}&\pmb{0.7922 (0.6103)} \\
                &2&      &0.8268 (0.5625)\\
                \cline{2-4}
                &1&\multirow{2}{*}{ZnSe}	 &\pmb{0.7607 (0.6491)}\\
                &2&		 &0.7861 (0.6180)\\
\hline
\multirow{4}{*}{\snmf}   &1&\multirow{2}{*}{\ch{BaTiO3}}&\pmb{0.0960 (0.9961)}\\	
                &2&		 &0.9732 (0.2299)\\
                \cline{2-4}
                &1&\multirow{2}{*}{ZnSe}  &0.9725 (0.2333)\\
                &2&		 &\pmb{0.1237 (0.9935)}\\
                \hline
    \end{tabular}
    \end{center}
    \label{tab:simuPDF}
\end{table}
Unlike the previous figures we just plot the agreement of the extracted component and the ground-truth curve that shows the best agreement.
The poor performance of the conventional 
NMF is evident in Fig.~\ref{fig:simuPDF_compare_error}(a) and (b), whereas again, even for this much smaller stretch, the \snmf algorithm still gives a good extraction of the physical components (Fig.~\ref{fig:simuPDF_compare_error}(c) and (d)).

\begin{figure}
  	\begin{center}
    \includegraphics[scale=0.55]{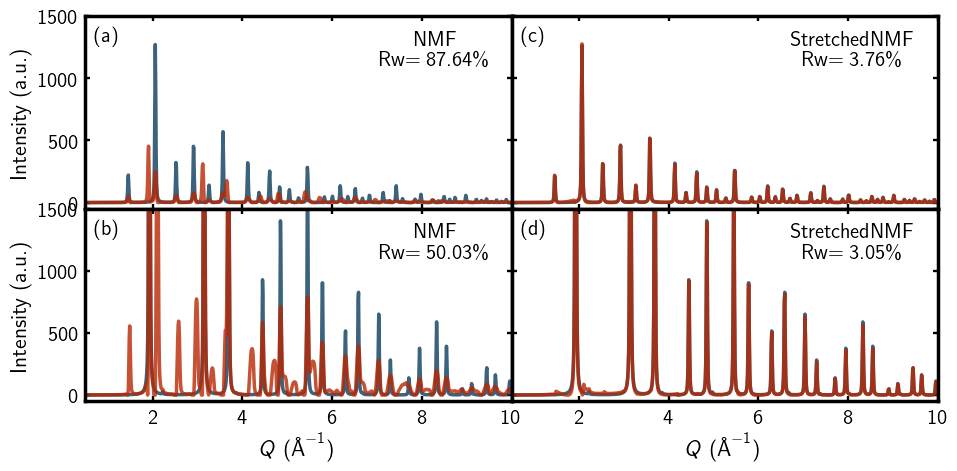}
  	\caption{Extracted components compared to the ground-truth components for powder PXRD patterns computed from the structures of \ch{BaTiO3} (blue curve in (a) and (c)) and ZnSe (blue curve in (b) and (d)). Extracted signals done with conventional NMF (red curves in (a) and (b)) and \snmf (red curves in (c) and (d)).} 	\label{fig:simuXRD_compare_noerror_strain02}
  	\end{center}
\end{figure}

We get the same overall result for the test on simulated PXRD data as for the PDF data.
The results are shown in Fig.~\ref{fig:simuXRD_compare_noerror_strain02} and Tab.~\ref{tab:simuXRD02}.
Again, \snmf gives a very good extraction of the physical components even for this small relative expansion coefficient (Fig.~\ref{fig:simuXRD_compare_noerror_strain02}(c) and (d)) whereas conventional NMF does not (Fig.~\ref{fig:simuXRD_compare_noerror_strain02}(a) and (b))
\begin{table}[]
\begin{center}
    \caption{Results of the comparison between the NMF extracted components and the ground-truth PXRDs on simulated PXRD data test with 2\% expansions on ZnSe. $R_w$ and PC are the residual and the Pearson Correlation values between the numerical solution and the ground truth. }
    \begin{tabular}{|c|c|c|c|}
    \hline
Method & Component &	Ground Truth	& $R_w$ (PC) \\
\hline
\multirow{4}{*}{Conventional NMF}&1&\multirow{2}{*}{\ch{BaTiO3}}&\pmb{0.8764 (0.4645)}\\
                &2&      &0.8869 (0.4449)\\
                \cline{2-4}
                &1&\multirow{2}{*}{ZnSe}	 &\pmb{0.5165 (0.8550)}\\
                &2&		 &0.5003 (0.8648)\\
\hline
\multirow{4}{*}{\snmf}   &1&\multirow{2}{*}{\ch{BaTiO3}}&\pmb{0.0376 (0.9993)}\\	
                &2&		 &0.9896 (0.1271)\\
                \cline{2-4}
                &1&\multirow{2}{*}{ZnSe}  &0.9991 (0.0217)\\
                &2&		 &\pmb{0.0305 (0.9995)}\\
                \hline
    \end{tabular}

    \label{tab:simuXRD02}
\end{center}
\end{table}

\begin{figure}
  	\begin{center}
   \includegraphics[scale=0.5]{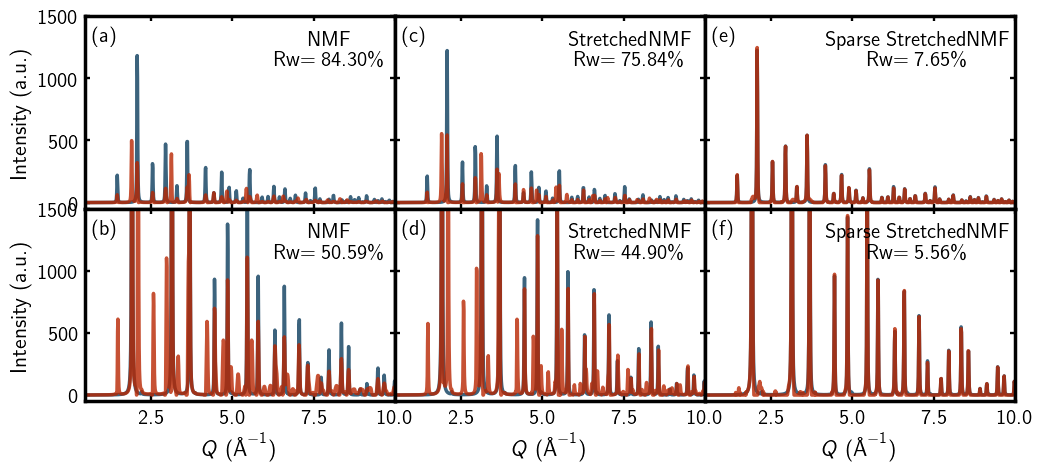}
  	\caption{Extracted components compared to the ground-truth components for powder PXRD patterns with 2\% and 1\% expansions
   , computed from the structures of \ch{BaTiO3} (blue curve in (a) (c) and (e)) and ZnSe (blue curve in (b), (d) and (f)). Extracted signals done with conventional NMF (red curves in (a) and (b)), \snmf (red curves in (c) and (d)) and \spsnmf (red curves in (e) and (f)).}
  	\label{fig:simuXRD_compare_noerror_strain01}
  	\end{center}
\end{figure}

The results are less ideal when the expansion rates are reduced further to \ch{BaTiO3} and ZnSe changing linearly from 1 to 1.02 and 1.01, respectively. The results are summarized in Tab.~\ref{tab:simuXRD01}, and Fig.~\ref{fig:simuXRD_compare_noerror_strain01}.
\begin{table}[]
    \centering
    \caption{Results of the comparison between the NMF extracted components and the ground-truth PXRDs on simulated PXRD data test with a 1\% differential expansion between the components. $R_w$ and PC are the residual and the Pearson Correlation values between the numerical solution and the ground truth.}
    \begin{tabular}{|c|c|c|c|}
    \hline
Method & Component &	Ground Truth	& $R_w$ (PC) \\
\hline
\multirow{4}{*}{Conventional NMF}&1&\multirow{2}{*}{\ch{BaTiO3}}&\pmb{0.8430 (0.5238)}\\
                &2&      &0.8491 (0.5142)\\
                \cline{2-4}
                &1&\multirow{2}{*}{ZnSe}	 &0.5248 (0.8493)\\
                &2&		 &\pmb{0.5059 (0.8609)}\\
\hline
\multirow{4}{*}{\snmf}   &1&\multirow{2}{*}{\ch{BaTiO3}}&\pmb{0.7584 (0.6413)}\\	
                &2&		 &0.8815 (0.4578)\\
                \cline{2-4}
                &1&\multirow{2}{*}{ZnSe}  &0.6482 (0.7574)\\
                &2&		 &\pmb{0.4490 (0.8921)}\\
                \hline
\multirow{4}{*}{\spsnmf} &1&\multirow{2}{*}{\ch{BaTiO3}}& \pmb{0.0765 (0.9971)}\\	
                &2&		 &0.9902 (0.1243)\\
                \cline{2-4}
                &1&\multirow{2}{*}{ZnSe}  &0.9951 (0.0823)\\
                &2&		 &\pmb{0.0556 (0.9985)}\\
                \hline
    \end{tabular}
    \label{tab:simuXRD01}
\end{table}

At this level of expansion, even the \snmf is not correctly extracting the physical components.
For example, it incorrectly assigns peaks in the spectrum of its extracted components in red at around $Q=1.5$, 2 and 2.5~\iaa (Fig.~\ref{fig:simuXRD_compare_noerror_strain01}(c) and (d)).  These same peaks are partially misassigned by the conventional NMF algorithm.

However, the \spsnmf algorithm does a good job of extracting physical components from the powder PXRD simulations (Fig.~\ref{fig:simuXRD_compare_noerror_strain01}(e) and (f)) even in this challenging case with a relatively small (1\%) differential expansion.
The components of \spsnmf are close to ground truths. This indicates that \spsnmf can enhance the performance of \snmf.

These tests show that the \snmf algorithm is able to extract physically meaningful PDF and PXRD signals from sets of data where the signals are unchanged except for a different relative stretch between the two curves.
If there is a large differential change in lattice parameter across the dataset \snmf can still extract ground-truth PDF and PXRD signals.
For relative stretches of a few percent, comparable to what might be expected for a mixture of compounds with a differential thermal expansion, this is also true for both PDF and PXRD data.
When the differential thermal expansion gets to around 1\%, \snmf starts to struggle to extract physical components.  However, for PXRD data the \spsnmf algorithm still performs well. We note that  the PDF data is not sparse, and therefore \spsnmf algorithm is applied only on PXRD data.

We should note that in these ground-truth tests on simulated data we wanted to test how well \snmf can handle datasets that contain stretches, for example, as might come from thermal expansion.
We therefore did not include in the simulation other effects of temperature changes such as increases in atomic displacement factors (ADPs).
In principle, we would like to develop a new algorithm that can eliminate changes in ADP in the same way as \snmf eliminates stretches.
This problem will be left for a future paper.
Preliminary tests on simulated data with combined stretching and increased-ADP effects indicate that \snmf and \spsnmf still perform reasonably well and clearly outperform the conventional NMF algorithm, but with larger errors than in the constant-ADP tests reported here.
Despite this known shortcoming, we would still like to see whether \snmf and \spsnmf can perform well on experimental data from a variable temperature experiment, and this is discussed in the following section.

\subsection{Results on measured PXRD data I}

Here, we test the NMF algorithms on measured PXRD data.
The data are from the \textit{in situ} chemical reaction experiment described in Section~\ref{sec:dataEXP1}.

Multi-component Rietveld refinements were carried out and indicate that the chemical components in this reaction are \ch{MgMn2O4}, orthorhombic \ch{YMnO3}, and rhombohedral and tetragonal \ch{YOCl} (\ch{rYOCl} and \ch{tYOCl}, respectively) where \ch{MgMn2O4} and \ch{tYOCl} are the dominant phases. The results of the Rietveld refinements for the two majority phases were used as ground truth against which to compare the performance of the NMF algorithms.

\begin{figure}
  	\begin{center}
  	\includegraphics[scale=0.5]{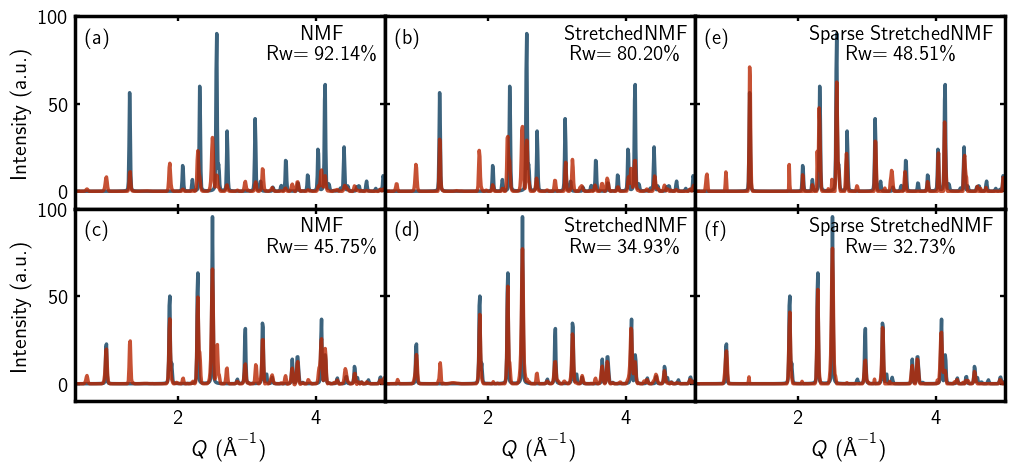}
  	\caption{Solutions on real PXRD data using conventional NMF, \snmf, and \spsnmf. The first row shows \ch{MgMn2O4} and the second row shows \ch{tYOCl}. The blue curves are the diffraction patterns obtained for those phases by multi-phase Rietveld refinement and are used as the desired component signal. The red curves represent the extracted NMF components.}
  	\label{fig:expXRD_compare_error}
  	\end{center}
\end{figure}

The results are shown in Fig.~\ref{fig:expXRD_compare_error} and the resulting $R_w$ and PC are listed in Tab.~\ref{tab:expXRD}.
\begin{table}[]
    \caption{Results of the comparison between the NMF extracted components and the ground truth from Rietveld refinement on real PXRD data test. $R_w$ and PC are the residual and the Pearson Correlation values between the numerical solution and the ground truth. }
    \centering
    \begin{tabular}{|c|c|c|c|}
    \hline
Method & Component &	Ground Truth	& $R_w$ (PC) \\
\hline
\multirow{4}{*}{Conventional NMF}&1&\multirow{2}{*}{\ch{MgMn2O4}}&0.9214(0.3402)\\
                &2&      &\pmb{0.9175(0.3519)}\\
                \cline{2-4}
                &1&\multirow{2}{*}{\ch{tYOCl}}	 &\pmb{0.4498(0.8863)}\\
                &2&		 &0.4575(0.8819)\\
\hline
\multirow{4}{*}{\snmf}   &1&\multirow{2}{*}{\ch{MgMn2O4}}&\pmb{0.8020(0.5655)} \\	
                &2&		 &0.9831(0.1246)\\
                \cline{2-4}
                &1&\multirow{2}{*}{\ch{tYOCl}}  &0.6433(0.7492) \\
                &2&		 &\pmb{0.3493(0.9330)}\\
                \hline
\multirow{4}{*}{\spsnmf}&1&\multirow{2}{*}{\ch{MgMn2O4}}&\pmb{0.4851(0.8682)}\\	
                &2&		 &0.9977(0.0137)\\
                \cline{2-4}
                &1&\multirow{2}{*}{\ch{tYOCl}}  &0.9646(0.2254)\\
                &2&		 &\pmb{0.3273(0.9422)}\\
                \hline
    \end{tabular}

    \label{tab:expXRD}
\end{table}
In Fig.~\ref{fig:expXRD_compare_error} the blue curves in the top row (a, b, e) are from the diffraction pattern of \ch{MgMn2O4} and the blue curves in the bottom row (c, d, f) are from t\ch{YOCl}.
The red curves in each panel show the relevant extracted component from the NMF algorithm used.
The columns are sorted by the NMF algorithm used.
The first column (a, c) used regular NMF, the second (b,d) used the \snmf algorithm, and the third column (e, f) used the \spsnmf algorithm.

All NMF solvers give reasonable results for the \ch{tYOCl} chemical component.
The peak positions are consistent with the ground truth, and the inconsistency of intensity is acceptable.
But for \ch{MgMn2O4}, the NMF and \snmf derived components are poor.
They are much better using the \spsnmf algorithm, which gives better agreement both visually and in terms of the \rw between the ground-truth and the extracted components.
For this case, from the perspective of separation ability, \spsnmf is superior to \snmf which is superior to the conventional NMF in this test.

\begin{figure}
  	\begin{center}
  	\includegraphics[scale=0.8]{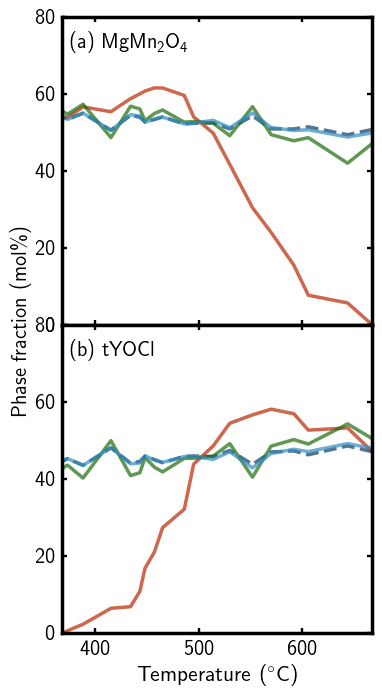}
  	\caption{The component weights vs. temperature of the data for (a) \ch{MgMn2O4} and (b) \ch{tYOCl}. The results obtained by a multi-phase Rietveld refinement can be considered as ground-truth and are shown as the dashed purple line.  The conventional NMF (red line) cannot capture the temperature evolution of the chemical components.  \snmf (green) does much better and \spsnmf (blue) yields almost exactly the Rietveld result.}
  	\label{fig:expXRD_weights}
  	\end{center}
\end{figure}

The scaled weights from all NMF solvers are compared to the weights from Rietveld refinement which can be considered as ground-truth. The results are shown in Fig.~\ref{fig:expXRD_weights}. 
The weights of the chemical components are not changing during the experiment and so we would expect the weights to be largely independent of temperature.   The conventional NMF clearly does not return constant weights and is getting confused by the thermal expansion in the data.
The \snmf and \spsnmf methods do yield almost constant weights.  
Rietveld refinements were carried out on these data-sets and can be  treated as a ground-truth.
The results of the Rietveld refinement are shown as the dashed curve.  
\snmf is doing quite well, but \spsnmf is doing very well in reproducing the results of the Rietveld refinement.

\subsection{Results on real PXRD data II}

We also tested the NMF algorithms on a real PXRD dataset from an in situ chemical reaction experiment, which was published in \cite{martinolichPolymorphSelectivitySuperconducting2015}. The data were measured as the temperature changed from 28 K to 370 K in 215 steps during the reaction
$$\ch{CuCl2} + \ch{Na2Se2}\longrightarrow \ch{CuSe2} + 2\ch{NaCl}, $$
where chemical components that found to appear during reaction are NaCl, CuSe, \ch{Cu2Se}, Se, pyrite, and marcasite.

\begin{figure}
  	\begin{center}
  	\includegraphics[scale=0.6]{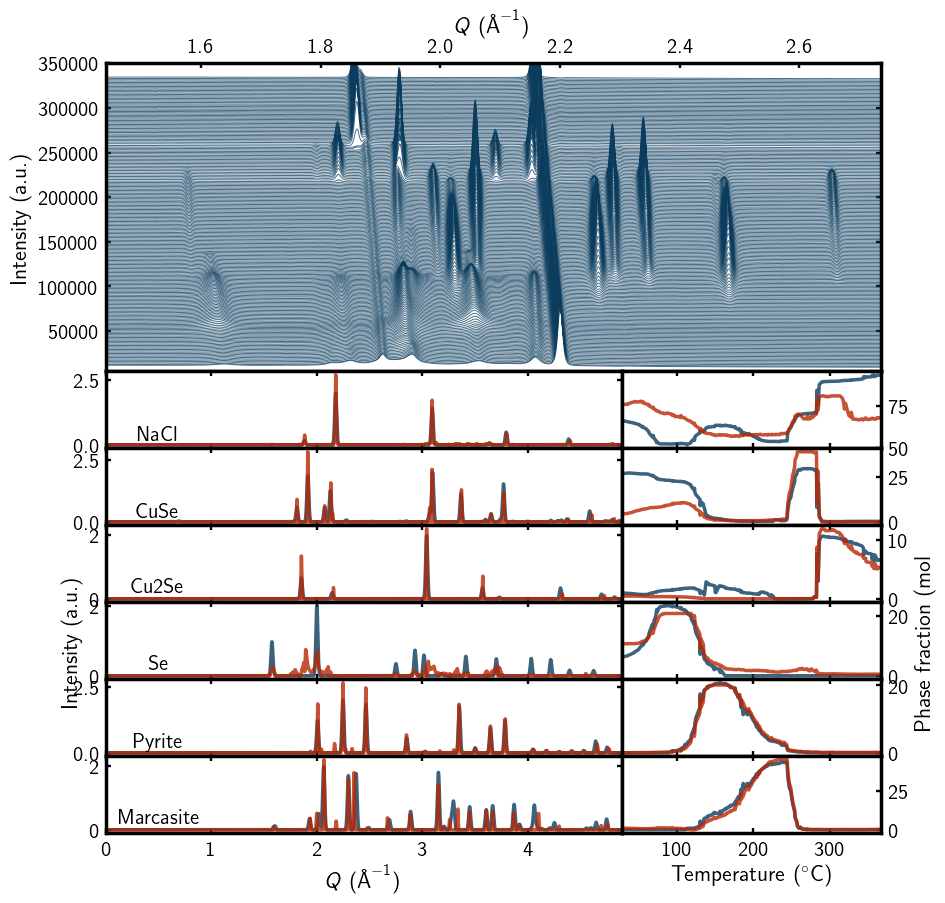}
  	\caption{The upper subplot shows the 215 experimental raw data PXRD curves, offset for clarity.
  The subplot below plots the ground-truth curves obtained from multi-phase Rietveld refinements \cite{martinolichPolymorphSelectivitySuperconducting2015} on the left and the correspondin Rietveld extracted weights in the right hand column, both in blue. The \snmf extracted components and weights are overlaid in the respective panel in red.}
  	\label{fig:expXRD_Jamie}
  	\end{center}
\end{figure}

The top panel in Fig.~\ref{fig:expXRD_Jamie} shows the measured PXRD data during the \textit{in situ} reaction experiment.
The curves obtained by a multi-phase Rietveld refinement fit \cite{martinolichPolymorphSelectivitySuperconducting2015} are shown in blue in the panels below.
The Rietveld refined phase weights are shown in blue in the right hand column below \cite{martinolichPolymorphSelectivitySuperconducting2015}.
The components extracted from a \snmf decomposition are shown in red, plotted on top of the ground-truth components, and the extracted weights are shown in red on top of the Rietveld extracted weights in the right hand column.
The results are very good and indicate that, except for \ch{Se}, the components obtained from \spsnmf matched well with the ground truth, as do the extracted weights.

This shows that the \snmf algorithm can be used as a rapid way to extract reliable components and weights from data collected at different temperatures.
This approach can be very helpful looking at large amounts of data very rapidly as it is being collected to look for known phases and unknown phases without having to carry out a complex multicomponent Rietveld campaign in real time.

\section{Discussion and Conclusion}\label{sec:5}

This paper presents a novel functional optimization model called \snmf, which is an extension to the traditional NMF model.
The proposed model introduces a new variable, the stretching factor, that enables the components to undergo stretching transformations.
Furthermore, a regularization term is incorporated to ensure the stretching factors are smooth over time or temperature.
To solve the optimization problem, we discretize it and employ Block Coordinate Descent (BCD) framework algorithms.
The initial experimental results indicate that for data where stretches in the signal are observed, such as diffraction data where thermal expansion has taken place, the proposed \snmf model outperforms the conventional NMF.
This is true even for PXRD and PDF data with small stretching degrees corresponding to realistic thermal expansivities.
However, a further enhancement to \snmf, which makes use of the sparsity of powder diffraction patterns, called \spsnmf allows correct extractions even for very small stretches where \snmf struggled.

Despite the utility described here, there are some limitations to the \snmf model.
One is where the stretching is anisotropic in a material.
This would require a model-dependent correction to account for different stretches in different crystallographic directions.

We also note that experimental noise can affect the outcome.
This has not been studied in detail in this paper, but we note that we obtained good results from real data that included noise.
To further address the noise issue, different regularization techniques can be utilized.

We note that the current model only considers stretching, adding shift transforms it into a first-order polynomial transformation.
In this case, only a new block is added to the computation, but a better approximation can be obtained.
Incorporating higher-order polynomial transformations could further balance the computational and approximative accuracy of the model.
Further research is needed to investigate and optimize the \snmf model's potential in overcoming these challenges.

Finally, we note that although the motivation for the development, and all the tests, were on diffraction data where underlying structures have undergone thermal expansion, the \snmf algorithm will work on any signal decomposition that smooth continuous variations in a stretching fact as a characteristic of the signal and it is not limited to use on diffraction data.




\ack{\paragraph{Acknowledgment}

We would like to thank Dr. Daniel Olds, for assistance during the measurements of the experimental PDF data.
The work described here was funded by the
Next Generation Synthesis Center (GENESIS), an Energy
Frontier Research Center funded by the U.S. Department
of Energy, Office of Science, Basic Energy Sciences
under Award Number DE-SC0019212.
X-ray PDF
measurements were conducted on beamline 28-ID-2 of the
National Synchrotron Light Source II, a US DOE Office of
Science User Facility operated for the DOE Office of Science
by Brookhaven National Laboratory under contract No. DESC0012704. Qiang Du is also partially supported by DOE-ASCRDE-SC0022317.
GEK received training and support as a part of QuADS: Quantitative Analysis of Dynamic Structures National Science Foundation Research Traineeship Program, grant number NSF DGE 1922639.}

\bibliographystyle{iucr}
\bibliography{rg_stretchednmf}

@article{liuUsingMachineLearning2019,
	title = {Using a machine learning approach to determine the space group of a structure from the atomic pair distribution function},
	volume = {75},
	copyright = {Copyright (c) 2019 International Union of Crystallography},
	issn = {2053-2733},
	url = {https://journals.iucr.org/a/issues/2019/04/00/ae5065/},
	doi = {10.1107/S2053273319005606},
	language = {en},
	number = {4},
	urldate = {2019-10-16},
	journal = {Acta Cryst A},
	author = {Liu, C.-H. and Tao, Y. and Hsu, D. and Du, Q. and Billinge, S. J. L.},
	month = jul,
	year = {2019},
	pages = {633--643},
}

@article{liuValidationNonnegativeMatrix2021a,
	title = {Validation of non-negative matrix factorization for rapid assessment of large sets of atomic pair distribution function data},
	volume = {54},
	copyright = {https://journals.iucr.org/copyright/licencetopublish.html},
	issn = {1600-5767},
	url = {https://journals.iucr.org/j/issues/2021/03/00/jl5015/},
	doi = {10.1107/S160057672100265X},
	language = {en},
	number = {3},
	urldate = {2021-05-20},
	journal = {J Appl Cryst},
	author = {Liu, C.-H. and Wright, C. J. and Gu, R. and Bandi, S. and Wustrow, A. and Todd, P. K. and O'Nolan, D. and Beauvais, M. L. and Neilson, J. R. and Chupas, P. J. and Chapman, K. W. and Billinge, S. J. L.},
	month = jun,
	year = {2021},
	note = {Number: 3
Publisher: International Union of Crystallography},
}

@article{yangCloudPlatformAtomic2021c,
	title = {A cloud platform for atomic pair distribution function analysis: {PDFitc}},
	volume = {77},
	copyright = {https://creativecommons.org/licenses/by/4.0/},
	issn = {2053-2733},
	shorttitle = {A cloud platform for atomic pair distribution function analysis},
	url = {https://journals.iucr.org/a/issues/2021/01/00/ae5091/},
	doi = {10.1107/S2053273320013066},
	language = {en},
	number = {1},
	urldate = {2020-12-15},
	journal = {Acta Crystallogr. A},
	author = {Yang, L. and Culbertson, E. A. and Thomas, N. K. and Vuong, H. T. and Kjær, E. T. S. and Jensen, K. M. Ø and Tucker, M. G. and Billinge, S. J. L.},
	month = jan,
	year = {2021},
	note = {Number: 1
Publisher: International Union of Crystallography},
}

@article{abdiPrincipalComponentAnalysis2010a,
	author = {Abdi, Herv{\'e} and Williams, Lynne J.},
	doi = {10.1002/wics.101},
	issn = {1939-0068},
	journal = {WIREs Computational Statistics},
	langid = {english},
	number = {4},
	pages = {433--459},
	title = {Principal Component Analysis},
	urldate = {2021-11-24},
	volume = {2},
	year = {2010}}

@article{berryAlgorithmsApplicationsApproximate2007b,
	author = {Berry, Michael W. and Browne, Murray and Langville, Amy N. and Pauca, V. Paul and Plemmons, Robert J.},
	doi = {10.1016/j.csda.2006.11.006},
	issn = {01679473},
	journal = {Computational Statistics \& Data Analysis},
	langid = {english},
	month = sep,
	number = {1},
	pages = {155--173},
	title = {Algorithms and Applications for Approximate Nonnegative Matrix Factorization},
	urldate = {2021-11-24},
	volume = {52},
	year = {2007}}

@article{buciuNonnegativeMatrixFactorization2008,
	author = {Buciu, Ioan and Nikolaidis, Nikos and Pitas, Ioannis},
	doi = {10.1109/TNN.2008.2000162},
	issn = {1941-0093},
	journal = {IEEE Transactions on Neural Networks},
	month = jun,
	number = {6},
	pages = {1090--1100},
	title = {Nonnegative {{Matrix Factorization}} in {{Polynomial Feature Space}}},
	volume = {19},
	year = {2008}}

@article{buciuNonnegativeMatrixFactorization2008a,
	author = {Buciu, Ioan},
	journal = {International Journal of Computers, Communications \& Control (IJCCC)},
	number = {3},
	pages = {67--74},
	title = {Nonnegative {{Matrix Factorization}}, {{A New Tool}} for {{Feature Extraction}}: {{Theory}} and {{Applications}}},
	volume = {3},
	year = {2008}}

@article{cichockiFastLocalAlgorithms2009,
	author = {Cichocki, Andrzej and Phan, Anh-Huy},
	doi = {10.1587/transfun.E92.A.708},
	issn = {0916-8508, 1745-1337},
	journal = {IEICE Transactions on Fundamentals of Electronics, Communications and Computer Sciences},
	langid = {english},
	number = {3},
	pages = {708--721},
	title = {Fast {{Local Algorithms}} for {{Large Scale Nonnegative Matrix}} and {{Tensor Factorizations}}},
	urldate = {2023-02-20},
	volume = {E92-A},
	year = {2009}}

@article{colemanInteriorTrustRegion1996,
	address = {{Philadelphia, United States}},
	author = {Coleman, Thomas F. and Li, Yuying},
	copyright = {[Copyright] \textcopyright{} 1996 \textcopyright{} Society for Industrial and Applied Mathematics},
	doi = {http://dx.doi.org/10.1137/0806023},
	issn = {10526234},
	journal = {SIAM Journal on Optimization},
	langid = {english},
	month = may,
	number = {2},
	pages = {28},
	publisher = {{Society for Industrial and Applied Mathematics}},
	title = {An {{Interior Trust Region Approach}} for {{Nonlinear Minimization Subject}} to {{Bounds}}},
	urldate = {2022-01-29},
	volume = {6},
	year = {1996}}

@book{dinnebierPowderDiffractionTheory2008a,
	author = {Dinnebier, Robert E. and Billinge, S. J. L.},
	googlebooks = {wmQ\_lFIMkFYC},
	isbn = {978-0-85404-231-9},
	langid = {english},
	publisher = {{Royal Society of Chemistry}},
	shorttitle = {Powder {{Diffraction}}},
	title = {Powder {{Diffraction}}: {{Theory}} and {{Practice}}},
	year = {2008}}

@book{egamiBraggPeaksStructural2012d,
	address = {{Amsterdam}},
	author = {Egami, T. and Billinge, S. J. L.},
	edition = {Second},
	isbn = {978-0-08-097133-9},
	number = {16},
	publisher = {{Elsevier}},
	series = {Pergamon Materials Series},
	title = {Underneath the {{Bragg}} Peaks: Structural Analysis of Complex Materials},
	year = {2012}}

@article{farrowRelationshipAtomicPair2009a,
	author = {Farrow, C. L. and Billinge, S. J. L.},
	copyright = {Copyright (c) 2009 International Union of Crystallography},
	doi = {10.1107/S0108767309009714},
	issn = {0108-7673},
	journal = {Acta Crystallographica Section A: Foundations of Crystallography},
	langid = {english},
	month = may,
	number = {3},
	pages = {232--239},
	shorttitle = {Relationship between the Atomic Pair Distribution Function and Small-Angle Scattering},
	title = {Relationship between the Atomic Pair Distribution Function and Small-Angle Scattering: Implications for Modeling of Nanoparticles},
	urldate = {2017-11-17},
	volume = {65},
	year = {2009}}

@inproceedings{gobinetApplicationNonnegativeMatrix2004,
	author = {Gobinet, Cyril and Perrin, Eric and Huez, R{\'e}gis},
	booktitle = {2004 12th {{European Signal Processing Conference}}},
	month = sep,
	pages = {1095--1098},
	title = {Application of {{Non-negative Matrix Factorization}} to Fluorescence Spectroscopy},
	year = {2004}}

@article{grippoConvergenceBlockNonlinear2000,
	author = {Grippo, L. and Sciandrone, M.},
	doi = {10.1016/S0167-6377(99)00074-7},
	issn = {01676377},
	journal = {Operations Research Letters},
	langid = {english},
	month = apr,
	number = {3},
	pages = {127--136},
	title = {On the Convergence of the Block Nonlinear {{Gauss}}\textendash{{Seidel}} Method under Convex Constraints},
	urldate = {2022-01-29},
	volume = {26},
	year = {2000}}

@article{guanNeNMFOptimalGradient2012,
	author = {Guan, Naiyang and Tao, Dacheng and Luo, Zhigang and Yuan, Bo},
	doi = {10.1109/TSP.2012.2190406},
	issn = {1941-0476},
	journal = {IEEE Transactions on Signal Processing},
	month = jun,
	number = {6},
	pages = {2882--2898},
	shorttitle = {{{NeNMF}}},
	title = {{{NeNMF}}: {{An Optimal Gradient Method}} for {{Nonnegative Matrix Factorization}}},
	volume = {60},
	year = {2012}}

@article{huangQuadraticRegularizationProjected2015,
	author = {Huang, Yakui and Liu, Hongwei and Zhou, Shuisheng},
	doi = {10.1007/s10618-014-0390-x},
	issn = {1384-5810, 1573-756X},
	journal = {Data Mining and Knowledge Discovery},
	langid = {english},
	month = nov,
	number = {6},
	pages = {1665--1684},
	title = {Quadratic Regularization Projected {{Barzilai}}\textendash{{Borwein}} Method for Nonnegative Matrix Factorization},
	urldate = {2022-01-29},
	volume = {29},
	year = {2015}}

@article{huaNonequilibriumMetalOxides2021,
	author = {Hua, Xiao and Allan, Phoebe K. and Gong, Chen and Chater, Philip A. and Schmidt, Ella M. and Geddes, Harry S. and Robertson, Alex W. and Bruce, Peter G. and Goodwin, Andrew L.},
	doi = {10.1038/s41467-020-20736-6},
	issn = {2041-1723},
	journal = {Nature Communications},
	langid = {english},
	month = jan,
	number = {1},
	pages = {561},
	title = {Non-Equilibrium Metal Oxides via Reconversion Chemistry in Lithium-Ion Batteries},
	urldate = {2023-03-01},
	volume = {12},
	year = {2021}}

@book{jolliffePrincipalComponentAnalysis2002,
	address = {{New York}},
	author = {Jolliffe, I. T.},
	edition = {2nd ed},
	isbn = {978-0-387-95442-4},
	lccn = {QA278.5 .J65 2002},
	publisher = {{Springer}},
	series = {Springer Series in Statistics},
	title = {Principal Component Analysis},
	year = {2002}}

@article{kimNonnegativeMatrixFactorization2008,
	address = {{Philadelphia, United States}},
	author = {Kim, Hyunsoo and Park, Haesun},
	copyright = {[Copyright] \textcopyright{} 2008 Society for Industrial and Applied Mathematics},
	doi = {http://dx.doi.org/10.1137/07069239X},
	issn = {08954798},
	journal = {SIAM Journal on Matrix Analysis and Applications},
	langid = {english},
	number = {2},
	pages = {18},
	publisher = {{Society for Industrial and Applied Mathematics}},
	title = {Nonnegative {{Matrix Factorization Based}} on {{Alternating Nonnegativity Constrained Least Squares}} and {{Active Set Method}}},
	urldate = {2022-01-29},
	volume = {30},
	year = {2008}}

@article{kusneHighthroughputDeterminationStructural2015a,
	author = {Kusne, A G and Keller, D and Anderson, A and Zaban, A and Takeuchi, I},
	doi = {10.1088/0957-4484/26/44/444002},
	issn = {0957-4484, 1361-6528},
	journal = {Nanotechnology},
	langid = {english},
	month = nov,
	number = {44},
	pages = {444002},
	title = {High-Throughput Determination of Structural Phase Diagram and Constituent Phases Using {{GRENDEL}}},
	urldate = {2022-01-29},
	volume = {26},
	year = {2015}}

@article{leeLearningPartsObjects1999d,
	author = {Lee, Daniel D. and Seung, H. Sebastian},
	doi = {10.1038/44565},
	issn = {0028-0836, 1476-4687},
	journal = {Nature},
	langid = {english},
	month = oct,
	number = {6755},
	pages = {788--791},
	title = {Learning the Parts of Objects by Non-Negative Matrix Factorization},
	urldate = {2021-11-24},
	volume = {401},
	year = {1999}}

@article{linProjectedGradientMethods2007,
	author = {Lin, Chih-Jen},
	doi = {10.1162/neco.2007.19.10.2756},
	issn = {0899-7667, 1530-888X},
	journal = {Neural Computation},
	langid = {english},
	month = oct,
	number = {10},
	pages = {2756--2779},
	title = {Projected {{Gradient Methods}} for {{Nonnegative Matrix Factorization}}},
	urldate = {2022-01-29},
	volume = {19},
	year = {2007}}

@article{longRapidIdentificationStructural2009a,
	author = {Long, C. J. and Bunker, D. and Li, X. and Karen, V. L. and Takeuchi, I.},
	doi = {10.1063/1.3216809},
	issn = {0034-6748, 1089-7623},
	journal = {Review of Scientific Instruments},
	langid = {english},
	month = oct,
	number = {10},
	pages = {103902},
	title = {Rapid Identification of Structural Phases in Combinatorial Thin-Film Libraries Using x-Ray Diffraction and Non-Negative Matrix Factorization},
	urldate = {2022-01-29},
	volume = {80},
	year = {2009}}

@article{martinolichPolymorphSelectivitySuperconducting2015,
	author = {Martinolich, Andrew J. and Kurzman, Joshua A. and Neilson, James R.},
	doi = {10.1021/ja512520z},
	issn = {0002-7863, 1520-5126},
	journal = {Journal of the American Chemical Society},
	langid = {english},
	month = mar,
	number = {11},
	pages = {3827--3833},
	title = {Polymorph {{Selectivity}} of {{Superconducting CuSe}} {\textsubscript{2}} {{Through Kinetic Control}} of {{Solid-State Metathesis}}},
	urldate = {2023-04-24},
	volume = {137},
	year = {2015}}

@inproceedings{morupShiftedNonNegativeMatrix2007,
	author = {Morup, Morten and Madsen, Kristoffer H. and Hansen, Lars K.},
	booktitle = {2007 {{IEEE Workshop}} on {{Machine Learning}} for {{Signal Processing}}},
	doi = {10.1109/MLSP.2007.4414296},
	issn = {2378-928X},
	month = aug,
	pages = {139--144},
	title = {Shifted {{Non-Negative Matrix Factorization}}},
	year = {2007}}

@article{paateroPositiveMatrixFactorization1994,
	author = {Paatero, Pentti and Tapper, Unto},
	doi = {10.1002/env.3170050203},
	issn = {11804009, 1099095X},
	journal = {Environmetrics},
	langid = {english},
	month = jun,
	number = {2},
	pages = {111--126},
	shorttitle = {Positive Matrix Factorization},
	title = {Positive Matrix Factorization: {{A}} Non-Negative Factor Model with Optimal Utilization of Error Estimates of Data Values},
	urldate = {2022-01-26},
	volume = {5},
	year = {1994}}

@book{pecharskyFundamentalsPowderDiffraction2008,
	author = {Pecharsky, Vitalij and Zavalij, Peter},
	isbn = {978-0-387-09579-0},
	langid = {english},
	month = nov,
	publisher = {{Springer Science \& Business Media}},
	title = {Fundamentals of {{Powder Diffraction}} and {{Structural Characterization}} of {{Materials}}, {{Second Edition}}},
	year = {2008}}

@article{rakit;am23,
	author = {Rakita, Yevgeny and Hart, James L. and Das, Partha Pratim and Shahrezaei, Sina and Foley, Daniel L. and Mathaudhu, Suveen Nigel and Nicolopoulos, Stavros and Taheri, Mitra L. and Billinge, Simon J. L.},
	doi = {10.1016/j.actamat.2022.118426},
	issn = {1359-6454},
	journal = {Acta Materialia},
	langid = {english},
	month = jan,
	pages = {118426},
	title = {Mapping Structural Heterogeneity at the Nanoscale with Scanning Nano-Structure Electron Microscopy ({{SNEM}})},
	urldate = {2023-01-21},
	volume = {242},
	year = {2023}}

@article{rakitaActiveReactionControl2020c,
	author = {Rakita, Yevgeny and O'Nolan, Daniel and McAuliffe, Rebecca D. and Veith, Gabriel M. and Chupas, Peter J. and Billinge, Simon J. L. and Chapman, Karena W.},
	doi = {10.1021/jacs.0c09418},
	issn = {0002-7863},
	journal = {Journal of the American Chemical Society},
	month = nov,
	number = {44},
	pages = {18758--18762},
	publisher = {{American Chemical Society}},
	title = {Active {{Reaction Control}} of {{Cu Redox State Based}} on {{Real-Time Feedback}} from {{In Situ Synchrotron Measurements}}},
	urldate = {2022-11-20},
	volume = {142},
	year = {2020}}

@article{renNonnegativeMatrixFactorization2018,
	author = {Ren, Bin and Pueyo, Laurent and Zhu, Guangtun Ben and Debes, John and Duch{\^e}ne, Gaspard},
	doi = {10.3847/1538-4357/aaa1f2},
	issn = {1538-4357},
	journal = {The Astrophysical Journal},
	langid = {english},
	month = jan,
	number = {2},
	pages = {104},
	shorttitle = {Non-Negative {{Matrix Factorization}}},
	title = {Non-Negative {{Matrix Factorization}}: {{Robust Extraction}} of {{Extended Structures}}},
	urldate = {2021-11-24},
	volume = {852},
	year = {2018}}

@techreport{sraNonnegativeMatrixApproximation2006,
	author = {Sra, Suvrit and Dhillon, Inderjit},
	institution = {{Computer Science Department, University of Texas at Austin}},
	title = {Nonnegative Matrix Approximation: {{Algorithms}} and Applications},
	year = {2006}}

@article{thatc;aca22,
	author = {Thatcher, Z. and Liu, C.-H. and Yang, L. and McBride, B. C. and Thinh Tran, G. and Wustrow, A. and Karlsen, M. A. and Neilson, J. R. and Ravnsb{\ae}k, D. B. and Billinge, S. J. L.},
	copyright = {https://journals.iucr.org/copyright/licencetopublish.html},
	doi = {10.1107/S2053273322002522},
	issn = {2053-2733},
	journal = {Acta Crystallographica Section A: Foundations and Advances},
	langid = {english},
	month = may,
	number = {3},
	publisher = {{International Union of Crystallography}},
	shorttitle = {{{nmfMapping}}},
	title = {{{nmfMapping}}: A Cloud-Based Web Application for Non-Negative Matrix Factorization of Powder Diffraction and Pair Distribution Function Datasets},
	urldate = {2022-04-06},
	volume = {78},
	year = {2022}}

@article{wangNonnegativeMatrixFactorization2013a,
	author = {Wang, Yu-Xiong and Zhang, Yu-Jin},
	doi = {10.1109/TKDE.2012.51},
	issn = {1041-4347},
	journal = {IEEE Transactions on Knowledge and Data Engineering},
	langid = {english},
	month = jun,
	number = {6},
	pages = {1336--1353},
	shorttitle = {Nonnegative {{Matrix Factorization}}},
	title = {Nonnegative {{Matrix Factorization}}: {{A Comprehensive Review}}},
	urldate = {2021-11-24},
	volume = {25},
	year = {2013}}

@article{xuGloballyConvergentAlgorithm2017,
	author = {Xu, Yangyang and Yin, Wotao},
	doi = {10.1007/s10915-017-0376-0},
	issn = {0885-7474, 1573-7691},
	journal = {Journal of Scientific Computing},
	langid = {english},
	month = aug,
	number = {2},
	pages = {700--734},
	title = {A {{Globally Convergent Algorithm}} for {{Nonconvex Optimization Based}} on {{Block Coordinate Update}}},
	urldate = {2022-01-29},
	volume = {72},
	year = {2017}}

@article{xuL1Regularization2010,
	author = {Xu, Zongben and Zhang, Hai and Wang, Yao and Chang, Xiangyu and Liang, Yong},
	doi = {10.1007/s11432-010-0090-0},
	journal = {Science China Information Sciences},
	langid = {english},
	month = jun,
	number = {6},
	pages = {1159--1169},
	title = {L1/2 Regularization},
	urldate = {2022-07-25},
	volume = {53},
	year = {2010}}

@article{yang;arxiv14,
	archiveprefix = {arxiv},
	author = {Yang, Xiaohao and Juhas, Pavol and Farrow, Christopher L. and Billinge, Simon J. L.},
	eprint = {1402.3163},
	journal = {arXiv:1402.3163 [cond-mat]},
	month = feb,
	primaryclass = {cond-mat},
	shorttitle = {{{xPDFsuite}}},
	title = {{{xPDFsuite}}: An End-to-End Software Solution for High Throughput Pair Distribution Function Transformation, Visualization and Analysis},
	urldate = {2021-11-23},
	year = {2015}}

@article{yangStructureminingScreeningStructure2020g,
	author = {Yang, L. and Juh{\'a}s, P. and Terban, M. W. and Tucker, M. G. and Billinge, S. J. L.},
	copyright = {https://creativecommons.org/licenses/by/4.0/},
	doi = {10.1107/S2053273320002028},
	issn = {2053-2733},
	journal = {Acta Crystallographica Section A: Foundations and Advances},
	langid = {english},
	month = may,
	number = {3},
	pages = {395--409},
	shorttitle = {Structure-Mining},
	title = {Structure-Mining: Screening Structure Models by Automated Fitting to the Atomic Pair Distribution Function over Large Numbers of Models},
	urldate = {2020-05-04},
	volume = {76},
	year = {2020}}

@article{su_thermal_2009,
	title = {Thermal expansion coefficient of {ZnSe} crystal between 17 and 1080 °C by interferometry},
	volume = {63},
	issn = {0167-577X},
	url = {https://www.sciencedirect.com/science/article/pii/S0167577X09002638},
	doi = {10.1016/j.matlet.2009.03.050},
	pages = {1475--1477},
	number = {17},
	journaltitle = {Materials Letters},
	shortjournal = {Materials Letters},
    journal = {Materials Letters},
	author = {Su, Ching-Hua and Feth, Shari and Lehoczky, S. L.},
	urldate = {2023-05-27},
	year = {2009},
	langid = {english},
}

@article{bland_thermal_1959,
	title = {{THE} {THERMAL} {EXPANSION} {OF} {CUBIC} {BARIUM} {TITANATE} ({BaTiO}3) {FROM} 350 °C {TO} 1050 °C},
	volume = {37},
	issn = {0008-4204},
	url = {https://cdnsciencepub.com/doi/10.1139/p59-046},
	doi = {10.1139/p59-046},
	pages = {417--421},
	number = {4},
	journaltitle = {Canadian Journal of Physics},
    journal = {Canadian Journal of Physics},
	shortjournal = {Can. J. Phys.},
	author = {Bland, J. A.},
	urldate = {2023-05-27},
	year = {1959},
	note = {Publisher: {NRC} Research Press},
}

@article{andreev_synthesis_1995,
	title = {Synthesis and Some Properties of Single Crystals of the \{Zn\$\_\{x\}\$Cd\$\_\{1-x\}\$S\} and \{{ZnS}\$\_\{y\}\$Se\$\_\{1-y\}\$\} Solid Solutions},
	pages = {1079--1082},
	issue = {7},
    volume = {40},
	journaltitle = {(Russian) Journal of Inorganic Chemistry (translated from Zhurnal Neorganicheskoi Khimii)},
    journal = {(Russian) Journal of Inorganic Chemistry (translated from Zhurnal Neorganicheskoi Khimii)},
	author = {Andreev, A.A. and Bulanyi, M.F. and Hayward, S.A. and Mozharovsikii, L.A.},
	year = {1995},
}

@article{keler_reaction_1960,
	title = {Reaction of {BaCO}\$\_3\$ with {TiO}\$\_2\$ and {ZrO}\$\_2\$ during heating},
	pages = {322--325},
	number = {5},
	journaltitle = {(Russian) Journal of Inorganic Chemistry (translated from Zhurnal Neorganicheskoi Khimii)},
    journal = {(Russian) Journal of Inorganic Chemistry (translated from Zhurnal Neorganicheskoi Khimii)},
	author = {Keler, E.K. and Karpenko, N.B.},
	year = {1960},
}

@article{juhas_complex_2015,
	title = {Complex modeling: a strategy and software program for combining multiple information sources to solve ill posed structure and nanostructure inverse problems},
	volume = {71},
	rights = {Copyright (c) 2015 International Union of Crystallography},
	issn = {2053-2733},
	url = {http://scripts.iucr.org/cgi-bin/paper?ae5008},
	doi = {10.1107/S2053273315014473},
	shorttitle = {Complex modeling},
	pages = {562--568},
	number = {6},
	journaltitle = {Acta Crystallographica Section A: Foundations and Advances},
    journal = {Acta Crystallographica Section A: Foundations and Advances},
	shortjournal = {Acta Cryst A},
	author = {Juhás, P. and Farrow, C. and Yang, X. and Knox, K. and Billinge, S.},
	urldate = {2022-08-30},
	date = {2015-11-01},
    year = {2015},
	langid = {english},
	note = {Number: 6
Publisher: International Union of Crystallography},
}

@software{porter_danporterdans_diffraction_2020,
	title = {{DanPorter}/Dans\_Diffraction},
	url = {https://zenodo.org/record/3859501},
	publisher = {Zenodo},
	author = {Porter, Dan},
	urldate = {2022-08-30},
	date = {2020-05-27},
    year = {2020},
	doi = {10.5281/zenodo.3859501},
	note = {Language: eng},
}

@article{ashiotis_fast_2015,
	title = {The fast azimuthal integration Python library: {pyFAI}},
	volume = {48},
	issn = {1600-5767},
	url = {//scripts.iucr.org/cgi-bin/paper?fv5028},
	doi = {10.1107/S1600576715004306},
	shorttitle = {The fast azimuthal integration Python library},
	pages = {510--519},
	number = {2},
	journaltitle = {Journal of Applied Crystallography},
    journal = {Journal of Applied Crystallography},
	shortjournal = {J Appl Cryst},
	author = {Ashiotis, G. and Deschildre, A. and Nawaz, Z. and Wright, J. P. and Karkoulis, D. and Picca, F. E. and Kieffer, J.},
	urldate = {2023-05-27},
	date = {2015-04-01},
    year = {2015},
	langid = {english},
	note = {Publisher: International Union of Crystallography},
}

@article{juhas2013pdfgetx3,
  title={PDFgetX3: a rapid and highly automatable program for processing powder diffraction data into total scattering pair distribution functions},
  author={Juh{\'a}s, Pavol and Davis, Timur and Farrow, Christopher L and Billinge, Simon JL},
  journal={Journal of Applied Crystallography},
  volume={46},
  number={2},
  pages={560--566},
  year={2013},
  publisher={International Union of Crystallography}
}

@article{onolan_thermal-gradient_2020,
	title = {A thermal-gradient approach to variable-temperature measurements resolved in space},
	volume = {53},
	rights = {© International Union of Crystallography, 2020},
	issn = {1600-5767},
	url = {https://onlinelibrary.wiley.com/doi/abs/10.1107/S160057672000415X},
	doi = {10.1107/S160057672000415X},
	pages = {662--670},
	number = {3},
	journal = {Journal of Applied Crystallography},
	author = {O'Nolan, Daniel and Huang, Guanglong and Kamm, Gabrielle E. and Grenier, Antonin and Liu, Chia-Hao and Todd, Paul K. and Wustrow, Allison and Thinh Tran, Gia and Montiel, David and Neilson, James R and Billinge, Simon J. L. and Chupas, Peter J. and Thornton, Katsuyo S. and Chapman, Karena W.},
	urldate = {2023-07-17},
	year = {2020},
	note = {\_eprint: https://onlinelibrary.wiley.com/doi/pdf/10.1107/S160057672000415X},
}

@article{chenNodeDistortionTunable2023,
  title = {Node {{Distortion}} as a {{Tunable Mechanism}} for {{Negative Thermal Expansion}} in {{Metal}}\textendash{{Organic Frameworks}}},
  author = {Chen, Zhihengyu and Stroscio, Gautam D. and Liu, Jian and Lu, Zhiyong and Hupp, Joseph T. and Gagliardi, Laura and Chapman, Karena W.},
  year = {2023},
  month = jan,
  journal = {Journal of the American Chemical Society},
  volume = {145},
  number = {1},
  pages = {268--276},
  publisher = {{American Chemical Society}},
  issn = {0002-7863},
  doi = {10.1021/jacs.2c09877},
  urldate = {2023-07-27}
}

@article{beauvaisResolvingSinglelayerNanosheets2021,
  title = {Resolving {{Single-layer Nanosheets}} as {{Short-lived Intermediates}} in the {{Solution Synthesis}} of {{FeS}}},
  author = {Beauvais, Michelle L. and Chupas, Peter J. and O'Nolan, Daniel and Parise, John B. and Chapman, Karena W.},
  year = {2021},
  month = jun,
  journal = {ACS MATERIALS LETTERS},
  volume = {3},
  number = {6},
  pages = {698--703},
  publisher = {{Amer Chemical Soc}},
  address = {{Washington}},
  issn = {2639-4979},
  doi = {10.1021/acsmaterialslett.1c00193},
  urldate = {2023-07-27},
  langid = {english}
}

@article{onolanMultimodalAnalyticalToolkit2021,
  title = {A Multimodal Analytical Toolkit to Resolve Correlated Reaction Pathways: The Case of Nanoparticle Formation in Zeolites},
  shorttitle = {A Multimodal Analytical Toolkit to Resolve Correlated Reaction Pathways},
  author = {O'Nolan, Daniel and Zhao, Haiyan and Chen, Zhihengyu and Grenier, Antonin and Beauvais, Michelle L. and Newton, Mark A. and Nenoff, Tina M. and Chupas, Peter J. and Chapman, Karena W.},
  year = {2021},
  month = oct,
  journal = {Chemical Science},
  volume = {12},
  number = {41},
  pages = {13836--13847},
  publisher = {{The Royal Society of Chemistry}},
  issn = {2041-6539},
  doi = {10.1039/D1SC04232G},
  urldate = {2023-07-27},
  langid = {english}
}

@article{rayderUnveilingUnexpectedModulatorCO22023,
  title = {Unveiling {{Unexpected Modulator-CO2 Dynamics}} within a {{Zirconium Metal}}\textendash{{Organic Framework}}},
  author = {Rayder, Thomas M. and Formalik, Filip and Vornholt, Simon M. and Frank, Hilliary and Lee, Seryeong and Alzayer, Maytham and Chen, Zhihengyu and Sengupta, Debabrata and Islamoglu, Timur and Paesani, Francesco and Chapman, Karena W. and Snurr, Randall Q. and Farha, Omar K.},
  year = {2023},
  month = may,
  journal = {Journal of the American Chemical Society},
  volume = {145},
  number = {20},
  pages = {11195--11205},
  publisher = {{American Chemical Society}},
  issn = {0002-7863},
  doi = {10.1021/jacs.3c01146},
  urldate = {2023-07-27}
}

\newpage
\section{Supplementary Information}
\subsection{Simulated PDF}
\begin{figure}
\includegraphics[scale=0.5]{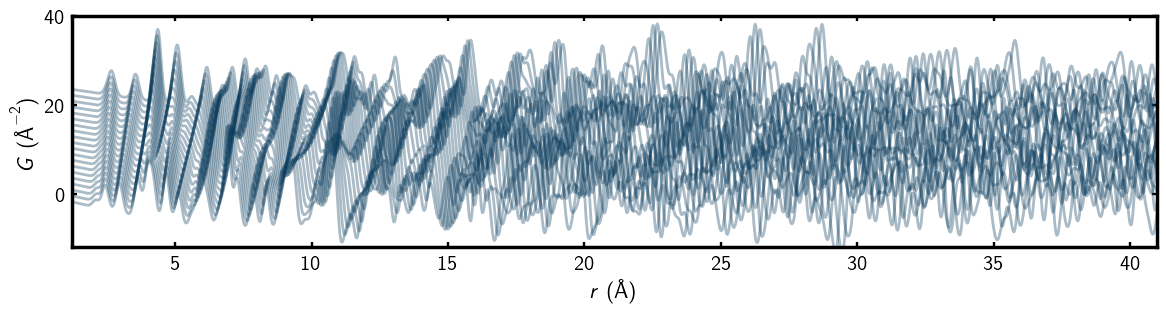}
  	\caption{The waterfall plot of the PDF signals used in the test. The figure is offset for clarity.}
\end{figure}

\subsection{Simulated PXRD}
\begin{figure}
\includegraphics[scale=0.5]{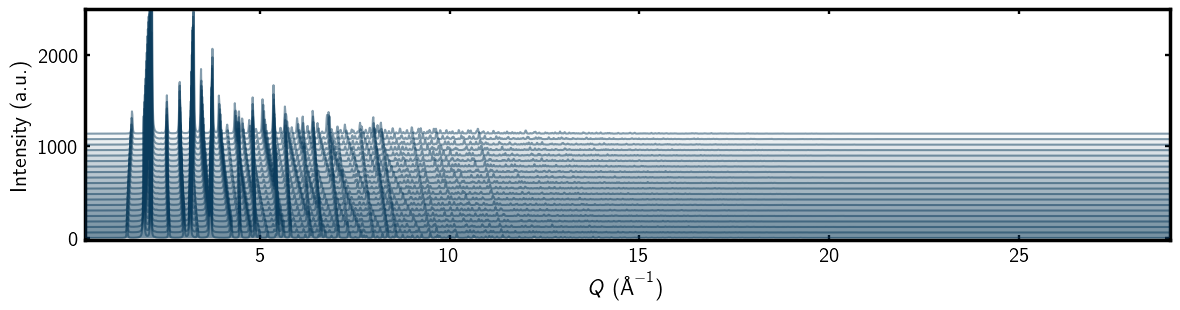}
  	\caption{The waterfall plot of the PXRD signals used in the test. The figure is offset for clarity.}
\end{figure}

\subsection{Simulated PDF and PXRD data with small expansion coefficients}

\begin{figure}
\includegraphics[scale=0.5]{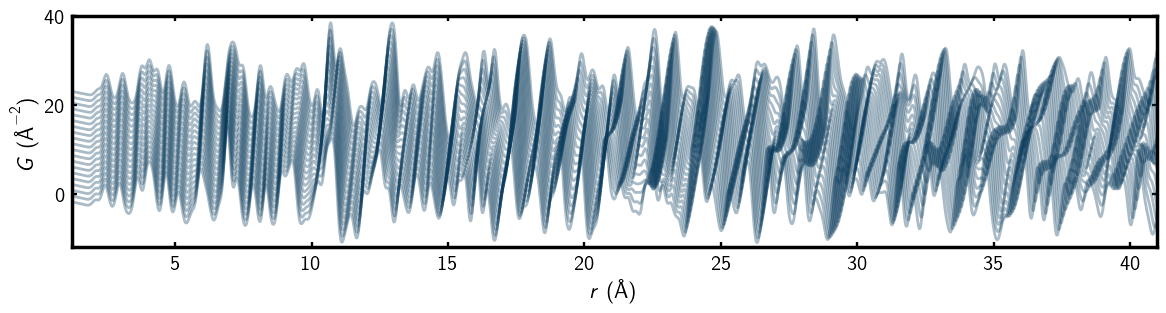}
  	\caption{The waterfall plot of the PDF signals used in the test. The figure is offset for clarity.}
\end{figure}

\begin{figure}
\includegraphics[scale=0.5]{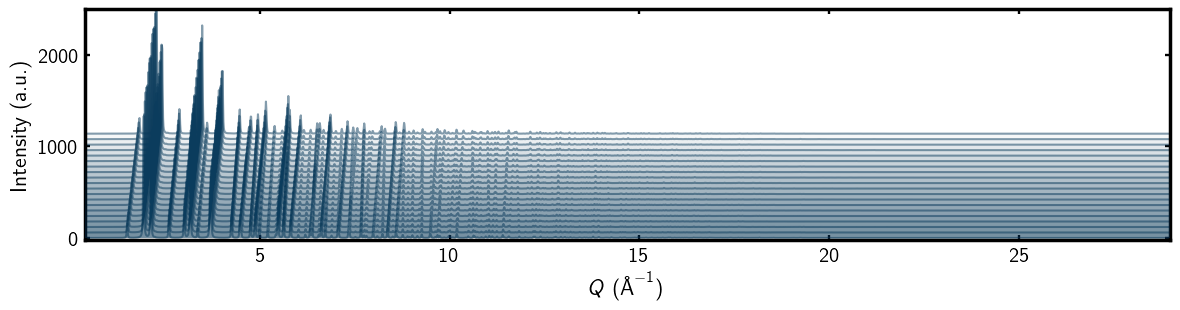}
  	\caption{The waterfall plot of the PXRD signals used in the test. The figure is offset for clarity.}
\end{figure}

\begin{figure}
\includegraphics[scale=0.5]{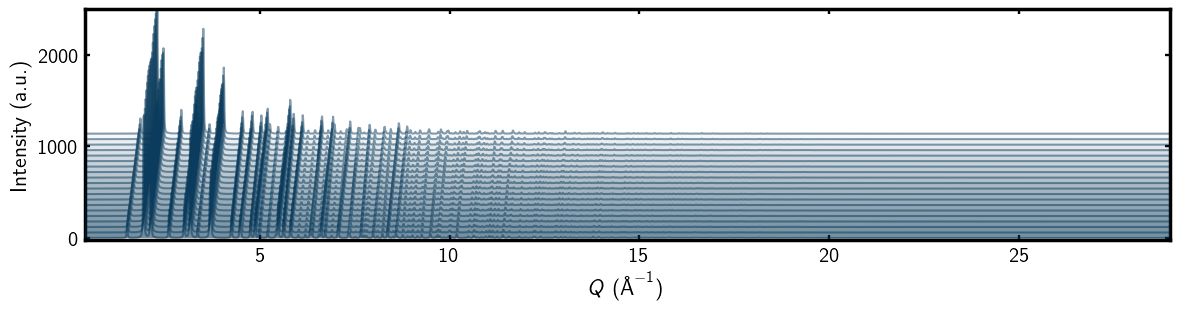}
  	\caption{The waterfall plot of the PXRD signals used in the test. The figure is offset for clarity.}
\end{figure}

\subsection{Real PXRD I}
\begin{figure}
\includegraphics[scale=0.5]{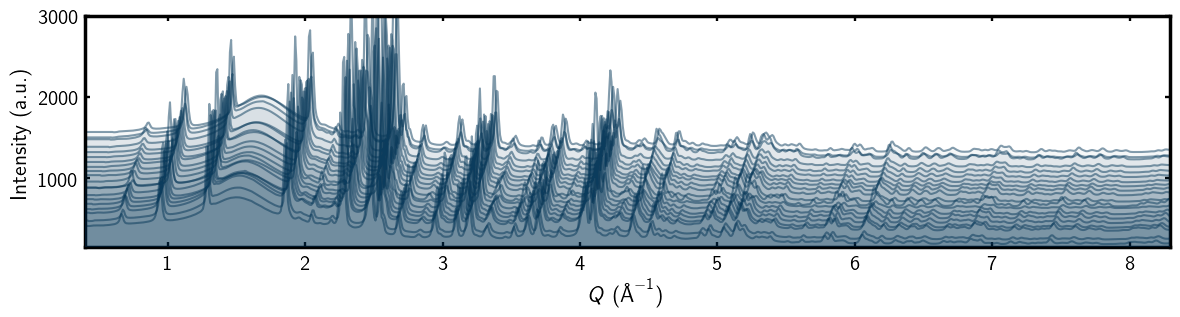}
  	\caption{The waterfall plot of the PXRD signals used in the test. The figure is offset for clarity.}
\end{figure}




\end{document}